\documentclass[twocolumn, showpacs, nofootinbib, aps, prd]{revtex4}

\usepackage{graphicx}
\usepackage{dcolumn}
\usepackage{color}
\usepackage{amsfonts}
\usepackage[colorlinks=true,linkcolor=blue,citecolor=blue, urlcolor=blue]{hyperref}

\begin{document}

\title{The Ratio $\mathcal{R}(D)$ and the $D$-meson Distribution Amplitude}

\author{Tao Zhong$^{1}$}
\email{zhongtao@htu.edu.cn}
\author{Yi Zhang$^{2}$}
\author{Xing-Gang Wu$^{2}$}
\email{wuxg@cqu.edu.cn}
\author{Hai-Bing Fu$^{3}$}
\author{Tao Huang$^4$}
\email{huangtao@ihep.ac.cn}

\address{$^1$ College of Physics and Materials Science, Henan Normal University, Xinxiang 453007, People's Republic of China\\
$^2$ Department of Physics, Chongqing University, Chongqing 401331, People's Republic of China \\
$^3$ School of Science, Guizhou Minzu University, Guiyang 550025, People's Republic of China \\
$^4$ Institute of High Energy Physics and Theoretical Physics Center for Science Facilities, Chinese Academy of Sciences, Beijing 100049, People's Republic of China}

\date{\today}

\begin{abstract}

In this paper, we calculate the $B\to D$ transition form factors (TFFs) within the light-cone sum rules (LCSR) and predict the ratio $\mathcal{R}(D)$. More accurate $D$-meson distribution amplitudes (DAs) are essential to get a more accurate theoretical prediction. We construct a new model for the twist-3 DAs $\phi^p_{3;D}$ and $\phi^\sigma_{3;D}$ based on the QCD sum rules under the background field theory for their moments as we have done for constructing the leading-twist DA $\phi_{2;D}$. As an application, we observe that the twist-3 contributions are sizable in whole $q^2$-region. Taking the twist-2 and twist-3 DAs into consideration, we obtain $f^{B\to D}_{+,0}(0) = 0.659^{+0.029}_{-0.032}$. As a combination of the Lattice QCD and the QCD LCSR predictions on the TFFs $f^{B\to D}_{+,0}(q^2)$, we predict $\mathcal{R}(D) = 0.320^{+0.018}_{-0.021}$, which improves is about $1.5\sigma$ deviation from the HFAG average of the Belle and BABAR data. At present the data are still of large errors, and we need further accurate measurements of the experiment to confirm whether there is signal of new physics from the ratio $\mathcal{R}(D)$.

\end{abstract}

\pacs{12.38.-t, 12.38.Bx, 14.40.Aq}

\maketitle

\section{introduction}

The $B$-meson physics provides a good platform for accurately testing the standard model (SM) and for finding the possible signal of new physics (NP), which has received much attention from physicists. In particular, the ratio $\mathcal{R}(D)$ in the semi-leptonic decay $B\to Dl\bar{\nu}_l$ has aroused people's great interests in recent years, since there sounds considerable difference between the experimental data and the SM theoretical predictions.

In year 2012, the BaBar Collaboration reports a first measurement on the ratio $\mathcal{R}(D)$, which is defined as
\begin{eqnarray}
\mathcal{R}(D) = \frac{\mathcal{B}(B\to D\tau\bar{\nu}_\tau)}{\mathcal{B}(B\to Dl^\prime\bar{\nu}_{l^\prime})}
\label{RD}
\end{eqnarray}
with $l^\prime$ stands for the light lepton $e$ or $\mu$. The BaBar Collaboration gives $\mathcal{R}^{\rm exp}(D)=0.440\pm 0.058 \pm 0.042$~\cite{BABAR_Lees:2012xj, BABAR_Lees:2013uzd}. The Belle collaboration gives a slightly smaller value $\mathcal{R}^{\rm exp}(D)=0.375 \pm 0.064 \pm 0.026$~\cite{BELLE_Huschle:2015rga}. The weighted average of those experimental measurements (HFAG average) gives $\mathcal{R}^{\rm exp}(D)=0.407\pm 0.039 \pm 0.024$~\cite{HFAG_Amhis:2014hma}. Many approaches have been tried to explain the data. Based on the heavy quark effective theory (HQET), Refs.\cite{HQET_Fajfer:2012vx, HQET_Tanaka:2010se} predict $\mathcal{R}(D)=0.302\pm 0.015$. By using the lattice QCD (LQCD), the FNAL/MILC Collaboration gives $\mathcal{R}(D)=0.299\pm 0.011$~\cite{LQCD_Lattice:2015rga} and the HPQCD Collaboration gives $\mathcal{R}(D)=0.300\pm 0.008$~\cite{LQCD_Na:2015kha}, whose average gives $\mathcal{R}(D)=0.300\pm 0.008$~\cite{Aoki:2016frl}. By using a global fit of the available LQCD predictions and experimental data, Ref.\cite{Bigi:2016mdz} predicts $\mathcal{R}(D)=0.299\pm 0.003$. Those SM predictions are consistent with each other within errors, however all of which are lower than its measured value, e.g. the LQCD prediction is about $2.1\sigma$ deviation from the the HFAG average. This inconsistency has motivated various speculations on the possible NP beyond the SM~\cite{NP_Celis:2012dk, NP_Celis:2013jha, NP_Li:2016vvp}.

Theoretical prediction on $\mathcal{R}(D)$ strongly depends on the $B\to D$ transition form factors (TFFs) $f_{+,0}^{B\to D}(q^2)$, which are mainly non-perturbative and can only be perturbatively calculated for large recoil region with $q^2\sim 0$. Thus before drawing definite conclusion, we have to know those TFFs better. The TFFs $f_{+,0}^{B\to D}(q^2)$ have been studied within the LQCD approach~\cite{LQCD_Lattice:2015rga, LQCD_Na:2015kha}, the pQCD factorization approach~\cite{PQCD_Fan:2015kna, PQCD_Fan:2013qz}, and the light-cone sum rules (LCSR) approach~\cite{Wang:2017jow, LCSR_Zuo:2006dk, Zuo:2006re, Fu:2013wqa, Zhang:2017rwz}. The pQCD approach is applicable for large recoil region and the LQCD approach is applicable for soft regions with large $q^2$. The LCSR approach involves both the hard and the soft contributions below $\sim 8 {\rm GeV}^{2}$. In the paper, we shall first adopt the LCSR approach to recalculate the TFFs and then combine the LQCD prediction to achieve a reliable prediction of the TFFs within the whole $q^2$-region.

The LCSRs for the TFFs $f_{+,0}^{B\to D}(q^2)$ can be expanded as a series over various $D$-meson light-cone distribution amplitudes (DAs). The high-twist DAs are generally power suppressed but could be sizable and helpful for a precise prediction. Several models for the leading-twist DA $\phi_{2;D}$ have been proposed in the literature~\cite{MODELI_Kurimoto:2002sb, MODELI_Keum:2003js, MODELII_Li:2008ts, MODELIII_Li:1999kna, MOLELIV_Guo:1991eb, MODELV_Grozin:1996pq, MODELVI_Kawamura:2001jm}. In Ref.\cite{Zhang:2017rwz}, we have studied the DA $\phi_{2;D}$ by recalculating its moments within the frame work of QCD SVZ sum rules~\cite{SVZ_Shifman:1978bx} under the background field theory (BFT)~\cite{Zhong:2014jla, Huang:1986wm, Huang:1989gv}. However at present, there is little research on the $D$-meson twist-3 DAs $\phi_{3;D}^p$ and $\phi_{3;D}^\sigma$. According to our experience, it is reasonable to assume that the twist-3 DAs shall have sizable contributions to the TFFs $f_{+,0}^{B\to D}(q^2)$. In previous pQCD treatment, the twist-3 DA $\phi_{3;D}^p$ is usually approximated by the leading-twist DA $\phi_{2;D}$ due to the difference between the moments of $\phi_{3;D}^p$ and $\phi_{2;D}$ is power suppressed by $\sim\mathcal{O}(\bar{\Lambda}/m_D)$ (where $\bar{\Lambda}=m_D-m_c$ with the $c$-quark mass $m_c$ and the $D$-meson mass), and the contribution from $\phi_{3;D}^\sigma$ is usually neglected which is suppressed by $\mathcal{O}(\bar{\Lambda}/m_D)$ compared to those of $\phi_{2;D}$ and $\phi_{3;D}^p$~\cite{Kurimoto:2002sb}. Thus more accurate twist-3 DAs shall also be helpful for achieving a precise prediction under pQCD factorization approach. In the paper, we will construct a new model for the $D$-meson twist-3 DAs $\phi_{3;D}^p$ and $\phi_{3;D}^\sigma$, whose moments will be determined by using the QCD SVZ sum rules under the BFT.

The remaining parts of the paper are organized as follows. The LCSRs for the TFFs $f_{+,0}^{B\to D}(q^2)$ with the next-to-leading order (NLO) corrections to the $D$-meson leading-twist DA contributions are given in Sec.II. The models for the $D$-meson DAs are discussed in Sec.III. A brief review of our previous model for the $D$-meson leading-twist DA $\phi_{2;D}$ is presented in Sec.III.A, which shall be improved by including the spin-space part into the wavefunctions. A new model for the twist-3 DAs $\phi_{3;D}^p$ and $\phi_{3;D}^\sigma$ is given in Sec.III.B. Numerical analysis and discussions are presented in Sec.IV. Sec.V is reserved for a summary.

\section{The Ratio $\mathcal{R}(D)$ and the $B\to D$ TFFs $f^{B\to D}_{+,0}(q^2)$ in the Light-Cone Sum Rules}

The ratio $\mathcal{R}(D)$ is determined by the branching ratio $\mathcal{B}(B\to D l\bar{\nu}_l)$, which can be calculated with
\begin{eqnarray}
\mathcal{B}(B\to Dl\bar{\nu}_l) = \tau_B \int^{(m_B - m_D)^2}_{m_l^2} dq^2 \frac{d\Gamma(B\to Dl\bar{\nu}_l)}{dq^2}
\label{Bra_Rat}
\end{eqnarray}
and
\begin{eqnarray} &&
\frac{d}{dq^2} \Gamma(B\to Dl\bar{\nu}_l) \nonumber\\
& =& \frac{G_F^2 |V_{\rm cb}|^2}{192\pi^3 m_B^3} \left( 1 - \frac{m_l^2}{q^2} \right)^2 \times\nonumber\\
&& \left[ \left( 1 + \frac{m_l^2}{2q^2} \right) \lambda^{3/2}(q^2) |f_+^{B\to D}(q^2)|^2 \right. \nonumber\\
&& + \left. \frac{3m_l^2}{2q^2} \left( m_B^2 - m_D^2 \right)^2 \lambda^{1/2}(q^2) |f_0^{B\to D}(q^2)|^2 \right],
\label{Dif_Dec_Wid}
\end{eqnarray}
where the phase-space factor $\lambda(q^2) = (m_B^2 + m_D^2 - q^2)^2 - 4m_B^2 m_D^2$, $\tau_B$ is the $B$-meson lifetimes, $m_{B}$ stands for the $B$-meson mass, $G_F$ is Fermi constant, $|V_{\rm cb}|$ is the CKM matrix element, and $m_l$ is the lepton mass.

The TFFs $f^{B\to D}_{+,0}(q^2)$ are important components of the ratio $\mathcal{R}(D)$, which are defined as
\begin{eqnarray}
\left<D(p)\left|\bar{c}\gamma_\mu b\right|B(p+q)\right> &=& 2f_{+}^{B\to D}(q^2) p_\mu + \left[ f_{+}^{B\to D}(q^2) \right. \nonumber\\
&& \left. +f_{-}^{B\to D}(q^2) \right] q_\mu
\label{f+}
\end{eqnarray}
and
\begin{eqnarray}
f_{0}^{B\to D}(q^2) = f_{+}^{B\to D}(q^2) + \frac{q^2}{m_B^2 - m_D^2} f_{-}^{B\to D}(q^2),
\label{f0}
\end{eqnarray}
where $p$ is the $D$-meson momentum and $q$ is the transition momentum. To determine the TFFs $f_{+,0}^{B\to D}(q^2)$, we adopt the LCSR method and take the correlator as
\begin{eqnarray}
\Pi_\mu (p,q)  &=& i \int d^4x e^{iq\cdot x} \nonumber\\
&& \times \left<D(p)\left| \bar{c}(x)\gamma_\mu b(x), m_b \bar{b}(0)i\gamma_5 q(0) \right|0\right>.
\label{cf}
\end{eqnarray}
Following the standard LCSR procedures, we obtain
\begin{eqnarray} &&
f^{B\to D}_+(q^2) \nonumber\\ &&
= \frac{e^{m_B^2/M^2}}{2m_B^2 f_B} \left[ F_0(q^2,M^2,s_0^B) + \frac{\alpha_s C_F}{4\pi} F_1(q^2,M^2,s_0^B) \right],
\label{srf+}
\end{eqnarray}
and
\begin{eqnarray} &&
f^{B\to D}_+(q^2) + f^{B\to D}_-(q^2) \nonumber\\ &&
= \frac{e^{m_B^2/M^2}}{m_B^2 f_B} \left[ \widetilde{F}_0(q^2,M^2,s_0^B) + \frac{\alpha_s C_F}{4\pi} \widetilde{F}_1(q^2,M^2,s_0^B) \right],
\label{srf+-}
\end{eqnarray}
where
\begin{eqnarray} &&
F_0(q^2,M^2,s_0^B) \nonumber\\ &&
= m_b^2 f_D \int^1_\Delta du \exp \left[ -\frac{m_b^2 - \bar{u}q^2 + u\bar{u}m_D^2}{uM^2} \right] \nonumber\\ &&
\times \left\{ \frac{\phi_{2;D}(u)}{u} + \frac{1}{m_b} \left[ \mu_D^p \phi_{3;D}^p(u) \right.\right. \nonumber\\ &&
+ \frac{\mu_D^\sigma}{6} \left( 1 - \frac{m_c^2}{m_D^2} \right) \left( \frac{2}{u} + \frac{4u m_b^2 m_D^2}{\left( m_b^2 - q^2 + u^2 m_D^2 \right)^2} \right. \nonumber\\ &&
- \left.\left.\left. \frac{m_b^2 + q^2 - u^2 m_D^2}{m_b^2 - q^2 + u^2 m_D^2} \frac{d}{du} \right) \phi_{3;D}^\sigma(u) \right] \right\},
\end{eqnarray}
\begin{eqnarray} &&
\widetilde{F}_0(q^2,M^2,s_0^B) \nonumber\\ &&
= m_b f_D \int^1_\Delta du \exp \left[ -\frac{m_b^2 - \bar{u}q^2 + u\bar{u}m_D^2}{uM^2} \right] \nonumber\\ &&
\times \left[ \mu_D^p \phi_{3;D}^p(u) + \frac{\mu_D^\sigma}{6u} \left( 1 - \frac{m_c^2}{m_D^2} \right) \frac{d\phi_{3;D}^\sigma(u)}{du} \right]
\end{eqnarray}
with
\begin{eqnarray}
\Delta &=& \left[ \sqrt{(s_0^B - q^2 - m_D^2)^2 + 4m_D^2 (m_b^2 - q^2)} \right. \nonumber\\
&& \left. -(s_0^B - q^2 - m_D^2) \right]/\left( 2m_D^2 \right).
\nonumber
\end{eqnarray}
The first terms in Eqs.(\ref{srf+}, \ref{srf+-}) are leading-order (LO) contributions for $f^{B\to D}_+(q^2)$ and $f^{B\to D}_+(q^2) + f^{B\to D}_-(q^2)$, respectively. $f_{B(D)}$ is the $B(D)$-meson decay constant, $m_{b}$ is the $b$-quark mass, $s_0^B$ is the threshold parameter, $M$ is the Borel parameter, $\mu_D^{p(\sigma)}$ is the normalization parameter of the DA $\phi_{3;D}^{p(\sigma)}$. The second terms in Eqs.(\ref{srf+}, \ref{srf+-}) are NLO corrections. Those LCSRs show that up to twist-3 accuracy, we have to know the twist-2 DA $\phi_{2;D}$ and twist-3 DAs $\phi_{3;D}^p$ and $\phi_{3;D}^\sigma$ well. There are also three-particle twist-3 terms, whose contributions are rather small and can be safely neglected. The $\bar{\Lambda}/m_D$ power-suppression and the $\alpha_s$-suppression are quantitatively at the same order level, thus in the paper, we shall consider the NLO corrections to the twist-2 terms and keep the twist-3 terms at the LO level. As an estimation, we neglect the charm quark current-mass effect to the twist-2 NLO terms of the $B\to D$ TFFs and take them as the same as the ones of the $B\to \pi$ TFFs~\cite{Duplancic:2008ix}.

\section{The $D$-meson leading-twist and twist-3 DAs}

\subsection{An improved model for the $D$-meson leading-twist DA $\phi_{2;D}$}

In Ref.\cite{Zhang:2017rwz} we have suggested a new light-cone harmonic oscillator model for the $D$-meson leading-twist wavefunction, which is based on the Brodsky-Huang-Lepage (BHL)-prescription~\cite{BHL1, BHL2, BHL3}, e.g.,
\begin{eqnarray}
\psi_{2:D}(x,\mathbf{k}_\perp) = \chi_{2:D}(x,\mathbf{k}_\perp) \psi^R_{2:D}(x,\mathbf{k}_\perp).
\label{wf}
\end{eqnarray}
In Eq.(\ref{wf}), $\chi_{2:D}(x,\mathbf{k}_\perp) = \tilde{m}/\sqrt{\mathbf{k}^2_\perp + \tilde{m}^2}$ with $\tilde{m} = \hat{m}_c x + \hat{m}_q(1-x)$ stands for the spin-space wavefunction. $\psi^R_{2:D}(x,\mathbf{k}_\perp)$ indicates the spatial wavefunction and takes the form
\begin{eqnarray}
\Psi^R_{2;D}(x,\mathbf{k}_\perp) &=& A_D \varphi_D(x) \nonumber\\
&\times& \exp \left[ -\frac{1}{\beta_D^2} \left( \frac{\mathbf{k}_\perp^2 + \hat{m}_c^2}{1-x} + \frac{\mathbf{k}_\perp^2 + \hat{m}_q^2}{x} \right) \right],
\label{wf_spatial}
\end{eqnarray}
with
\begin{eqnarray}
\varphi_D(x) = 1 + \sum^4_{n=1} B_n^D C_n^{3/2}(2x-1), \nonumber
\end{eqnarray}
and $\mathbf{k}_\perp$ is the transverse momentum, $\hat{m}_c$ and $\hat{m}_q$ are constituent charm-quark and light-quark masses, and we adopt $\hat{m}_c = 1.5 \rm GeV$ and $\hat{m}_q = 0.3 \rm GeV$. This model is applicable for both $\overline{D}^0$ and $D^-$ leading-twist wavefunctions since the mass difference between $u$ and $d$ is negligible. One can obtain the leading-twist wavefunction of $D^0$ or $D^+$ by replacing $x$ with $1-x$ in Eq.(\ref{wf}).

After integrating out the transverse momentum $\bf{k}_\perp$ component in wavefunction $\Psi_{2;D}(x,\mathbf{k}_\perp)$, the $D$-meson leading-twist DA $\phi_{2;D}$ can be obtained. We have approximately taken $\chi_{2;D}\to 1$ in our previous treatment~\cite{Zhang:2017rwz}; At present, we keep the $\chi_{2;D}$-terms to obtain a more accurate behavior for $\phi_{2;D}$, i.e.
\begin{eqnarray}
\phi_{2;D}(x,\mu_0) &=& \frac{\sqrt{3} A_D \tilde{m} \beta_D}{2 \pi^{3/2} f_D} \sqrt{x(1-x)} \varphi_D(x) \nonumber\\
&\times& \exp \left[ - \frac{\hat{m}_c^2x + \hat{m}_q^2(1-x) - \tilde{m}^2}{8\beta_D^2 x(1-x)} \right] \nonumber\\
&\times& \left\{ {\rm Erf} \left[ \sqrt{\frac{\tilde{m}^2 + \mu_0^2}{8\beta_D^2 x(1-x)}} \right] \right. \nonumber\\
&-& \left. {\rm Erf} \left[ \sqrt{\frac{\tilde{m}^2}{8\beta_D^2 x(1-x)}} \right] \right\},
\label{phi}
\end{eqnarray}
where $\mu_0$ is the factorization scale, ${\rm Erf}(x)$ is the error function. The input parameters $A_D$, $B_n^D$ and $\beta_D$ can be fixed by the normalization condition of $\phi_{2;D}$, the probability of finding the leading Fock-state $\left|\bar{c}q\right>$ in the $D$-meson Fock-state expansion which can be taken as $P_D \simeq 0.8~$\cite{MOLELIV_Guo:1991eb}, and the known moments $\left<\xi^n\right>_D$ [or the known Gegenbauer moments $a^D_n$] of $\phi_{2;D}$. Furthermore, the average value of the squared $D$-meson transverse momentum $\left<\bf{k}_\perp^2\right>_D$ can be calculated via the following way
\begin{eqnarray}
\left<\bf{k}_\perp^2\right>_D &=& \frac{1}{P_D} \int^1_0 dx \int \frac{d^2\mathbf{k}_\perp}{16\pi^3} \mathbf{k}_\perp^2 \left| \Psi^R_{2;D}(x,\mathbf{k}_\perp) \right|^2 \nonumber\\
&=& \frac{A_D^2 \beta_D^4}{\pi^2 P_D} \int^1_0 dx x^2(1-x)^2 (\varphi_D(x))^2 \nonumber\\
&\times& \exp \left[ -\frac{\hat{m}_c^2x + \hat{m}_q^2(1-x)}{4\beta_D^2 x(1-x)} \right],
\label{K2}
\end{eqnarray}
which can be used to constrain the behaviors of the $D$-meson twist-3 DAs $\phi_{3;D}^p$ and $\phi_{3;D}^\sigma$.

\subsection{A new model for the $D$-meson twist-3 DAs}

Following the above idea of constructing the $D$-meson leading-twist DA, we suggest the following model for the twist-3 DA $\phi_{3;D}^p$
\begin{eqnarray}
\phi_{3;D}^p(x,\mu_0) &=& \frac{\sqrt{3} A_D^p \tilde{m} \beta_D^p}{2 \pi^{3/2} f_D} \sqrt{x(1-x)} \varphi_D^p(x) \nonumber\\
&\times& \exp \left[ - \frac{\hat{m}_c^2x + \hat{m}_q^2(1-x) - \tilde{m}^2}{8(\beta_D^p)^2 x(1-x)} \right] \nonumber\\
&\times& \left\{ {\rm Erf} \left[ \sqrt{\frac{\tilde{m}^2 + \mu_0^2}{8(\beta_D^p)^2 x(1-x)}} \right] \right. \nonumber\\
&-& \left. {\rm Erf} \left[ \sqrt{\frac{\tilde{m}^2}{8(\beta_D^p)^2 x(1-x)}} \right] \right\},
\label{phipmodel}
\end{eqnarray}
with
\begin{eqnarray}
\varphi_D^p(x) = 1 + \sum_{n=1}^4 B_n^{D,p} \times C_n^{1/2}(2x-1).
\label{varphi}
\end{eqnarray}
The model parameters $A_D^p$, $B_n^{D,p}$ and $\beta_D^p$ are determined by the following constraints:
\begin{itemize}
\item The normalization condition of $\phi_{3;D}^p$,
\begin{eqnarray}
\int^1_0 dx \phi_{3;D}^p(x,\mu_0) = 1.
\label{NC}
\end{eqnarray}

\item The average value of the squared $D$ transverse momentum $\left<\bf{k}_\perp^2\right>_D$, i.e.
\begin{eqnarray}
\left<\bf{k}_\perp^2\right>_D &=& \frac{(A_D^p)^2 (\beta_D^p)^4}{\pi^2 P_D} \int^1_0 dx x^2(1-x)^2 (\varphi_D^p(x))^2 \nonumber\\
&\times& \exp \left[ -\frac{\hat{m}_c^2x + \hat{m}_q^2(1-x)}{4(\beta_D^p)^2 x(1-x)} \right],
\label{K}
\end{eqnarray}

\item The moments $\left<\xi^n_p\right>_D$ of the $D$-meson twist-3 DA $\phi_{3;D}^p$ are defined as
\begin{eqnarray}
\left<\xi^n_p\right>_D |_{\mu_0} = \int^1_0 dx (2x-1)^n \phi_{3;D}^p(x,\mu_0),
\label{xin}
\end{eqnarray}
which can be calculated by using the QCD sum rules under the framework of BFT.
\end{itemize}

The twist-3 DA $\phi_{3;D}^\sigma$ can be constructed under the same way. By replacing the upper index `$p$' with `$\sigma$' in Eq.(\ref{phipmodel}) and taking the expansion
\begin{eqnarray}
\varphi_D^\sigma(x) = 1 + \sum_{n=1}^4 B_n^{D,\sigma} \times C_n^{3/2}(2x-1),
\end{eqnarray}
we obtain the model for $\phi_{3;D}^\sigma$.

In above equations, the factorization scale is taken as $\mu_0\sim 1$ GeV, the DAs at any other scale can be obtained via the conventional evolution equation~\cite{Lepage:1980fj}.

In addition to the known parameters, our task left is to determine the moments of the twist-3 DAs $\phi_{3;D}^p$ and $\phi_{3;D}^\sigma$. We adopt the following correlators to achieve the sum rules for the moments $\left<\xi^n_p\right>_D$ and $\left<\xi^n_\sigma\right>_D$, i.e.
\begin{eqnarray}
\Pi^{p}_D(q) &=& i\int d^4x e^{iq\cdot x} \left<0\left| T\left\{ J^{\rm PS}_n(x) J^{\rm PS \dagger}_0 \right\} \right|0\right> \nonumber\\
&=& (z\cdot q)^n I^{p}_D(q^2)
\label{cfps}
\end{eqnarray}
and
\begin{eqnarray}
\Pi^{\sigma}_D(q) &=& i\int d^4x e^{iq\cdot x} \left<0\left| T\left\{ J^{\rm PT}_n(x) J^{\rm PS \dagger}_0 \right\} \right|0\right> \nonumber\\
&=& -i(q_\mu z_\nu - q_\nu z_\mu) (z\cdot q)^n I^{\sigma}_D(q^2),
\label{cfpt}
\end{eqnarray}
where $z^2 = 0$, $J^{\rm PS}_n(x)$ and $J^{\rm PT}_n(x)$ are pseudo-scalar and pseudo-tensor currents
\begin{eqnarray}
J^{\rm PS}_n(x) &=& \bar{c}(x)\gamma_5(iz\cdot \tensor{D})^n q(x), \label{jps} \\
J^{\rm PT}_n(x) &=& \bar{c}(x)\sigma_{\mu\nu}\gamma_5(iz\cdot \tensor{D})^{n+1} q(x) \label{jpt}
\end{eqnarray}
with $\sigma_{\mu\nu} = \frac{i}{2}(\gamma_\mu\gamma_\nu-\gamma_\nu\gamma_\mu)$.

Following the standard procedures of the SVZ QCD sum rules under the BFT~\cite{Zhang:2017rwz,Zhong:2014jla} with the help of the relations between the hadronic transition matrix elements and the moments
\begin{eqnarray}
\left<0\left| J^{\rm PS}_n(0) \right|D(q)\right> &=& -i\mu_D^p f_D \left<\xi^n_p\right>_D (z\cdot q)^n , \label{xip} \\
\left<0\left| J^{\rm PT}_n(0) \right|D(q)\right> &=& -\frac{n+1}{3} \mu_D^\sigma f_D \left( 1 - \frac{m_c^2}{m_D^2} \right) \left<\xi^n_\sigma\right>_D \nonumber\\
&\times& (q_\mu z_\nu - q_\nu z_\mu) (z\cdot q)^n, \label{xit}
\end{eqnarray}
one can obtain the required sum rules, i.e.
\begin{eqnarray}
\left<\xi^n_p\right>_D &=& \frac{M^2 e^{\frac{m_D^2}{M^2}}}{(\mu_D^p)^2 f_D^2} \left\{ \frac{1}{\pi} \frac{1}{M^2} \int^{s^D_0}_{m_c^2} ds e^{-\frac{s}{M^2}} {\rm Im} I^{p}_{D,\rm pert.} \right. \nonumber\\
&+& \hat{L}_M I^{p}_{D,\left<\bar{q}q\right>} + \hat{L}_M I^{p}_{D,\left<G^2\right>} + \hat{L}_M I^{p}_{D,\left<\bar{q}Gq\right>} \nonumber\\
&+& \left. \hat{L}_M I^{p}_{D,\left<\bar{q}q\right>^2} + \hat{L}_M I^{p}_{D,\left<G^3\right>} \right\}, \label{srxips} \\
\left<\xi^n_\sigma\right>_D &=& \frac{3M^2 e^{\frac{m_D^2}{M^2}}}{(n+1)\mu_D^p \mu_D^\sigma f_D^2} \frac{m_D^2}{m_D^2-m_c^2} \nonumber\\
&\times& \left\{ \frac{1}{\pi} \frac{1}{M^2} \int^{s^D_0}_{m_c^2} ds e^{-\frac{s}{M^2}} {\rm Im} I^{\sigma}_{D,\rm pert.} \right. \nonumber\\
&+& \hat{L}_M I^{\sigma}_{D,\left<\bar{q}q\right>} + \hat{L}_M I^{\sigma}_{D,\left<G^2\right>} + \hat{L}_M I^{\sigma}_{D,\left<\bar{q}Gq\right>} \nonumber\\
&+& \left. \hat{L}_M I^{\sigma}_{D,\left<\bar{q}q\right>^2} + \hat{L}_M I^{\sigma}_{D,\left<G^3\right>} \right\}, \label{srxipt}
\end{eqnarray}
where $\hat{L}_M$ is the Borel transformation operator. The explicit expressions for the short notations like ${\rm Im} I^{p}_{D,\rm pert.}$, $\hat{L}_M I^{p}_{D,\left<\bar{q}q\right>}$ and etc. are presented in the Appendix.

\section{numerical analysis}

\subsection{Input parameters}

To determine the moments of the $D$-meson twist-3 DAs, we take~\cite{PDG_Olive:2016xmw}
\begin{eqnarray}
m_{D^-} &=& 1869.59 \pm 0.09 {\rm MeV}, \nonumber\\
f_D &=& 203.7 \pm 4.7 \pm 0.6 {\rm MeV}, \nonumber\\
\bar{m}_{c}(\bar{m}_c) &=& 1.28 \pm 0.03 {\rm GeV}, \nonumber\\
\bar{m}_d(2{\rm GeV}) &=& 4.7^{+0.5}_{-0.4} {\rm MeV}.
\end{eqnarray}
For the condensates up to dimension-six, we take~\cite{SRREV_Colangelo:2000dp}
\begin{eqnarray}
\left<\bar{q}q\right>(1 {\rm GeV}) &=& -(240 \pm 10 {\rm MeV})^3, \nonumber\\
\left<g_s\bar{q}\sigma TGq\right>(1 {\rm GeV}) &=& 0.8 \left<\bar{q}q\right>(1 {\rm GeV}), \nonumber\\
\left<\alpha_sG^2\right> &=& 0.038 \pm 0.011 {\rm GeV}^4, \nonumber\\
\left<g_s^3fG^3\right> &=& 0.045 {\rm GeV}^6, \nonumber\\
\left<g_s\bar{q}q\right>^2 &=& 1.8 \times 10^{-3} {\rm GeV}^6.
\end{eqnarray}
The scale-dependent parameters at any other scales can be obtained by using the renormalization group equation~\cite{RGE_Yang:1993bp, RGE_Hwang:1994vp}. As exceptions, the gluon-condensates $\left<\alpha_sG^2\right>$ and $\left<g_s^3fG^3\right>$ are scale-independent, and we ignore the scale-dependence of the four-quark condensate $\left<g_s\bar{q}q\right>^2$, whose contribution to the twist-3 DA moment is small. In doing the calculation, we take the renormalization scale $\mu=M$, since the Borel parameter $M$ characterizes the typical momentum flow of the process. For the continuous threshold $s_0^{D}$, as discussed in Ref.\cite{Zhang:2017rwz}, we take $s_0^D \simeq 6.5 \textrm{GeV}^2$.

\subsection{Update for the $D$-meson twist-2 DA $\phi_{2;D}$}

\begin{table}[htb]
\caption{Criteria for determining the Borel windows of the moments $\left<\xi^{n=1,\cdots,4}\right>_D$. }
\begin{tabular}{ c  c  c }
\hline
~~ & ~Continue~ & ~Dimension-six~ \\
~~ & ~Contribution ($\%$)~ & ~Contribution ($\%$)~ \\
\hline
~$\left<\xi^1\right>_D$~ & ~$<15$~ & ~$<5$~ \\
~$\left<\xi^2\right>_D$~ & ~$<30$~ & ~$<10$~ \\
~$\left<\xi^3\right>_D$~ & ~$<30$~ & ~$<10$~ \\
~$\left<\xi^4\right>_D$~ & ~$<45$~ & ~$<15$~ \\
\hline
\end{tabular}
\label{t2criterion}
\end{table}

\begin{table}[htb]
\caption{The Borel windows and the allowable regions for the moments $\left<\xi^{n=1,\cdots,4}\right>_D$. All other input parameters are set to be their central values. }
\begin{tabular}{ c  c  c }
\hline
~~ & ~$M^2$~ & ~Value~ \\
\hline
~$\left<\xi^1\right>_D$~ & ~$[2.667,7.095]$~ & ~$[-0.433,-0.399]$~ \\
~$\left<\xi^2\right>_D$~ & ~$[2.627,3.374]$~ & ~$[0.319,0.321]$~ \\
~$\left<\xi^3\right>_D$~ & ~$[3.671,14.862]$~ & ~$[-0.192,-0.169]$~ \\
~$\left<\xi^4\right>_D$~ & ~$[3.589,5.257]$~ & ~$[0.157,0.148]$~ \\
\hline
\end{tabular}
\label{t2bw}
\end{table}

Here, we adopt the value of $\left<g_s^3fG^3\right>$ as the commonly used one suggested by Ref.\cite{SRREV_Colangelo:2000dp}, instead of the one adopted in our previous paper~\cite{Zhang:2017rwz}, then the corresponding results about the moments of the $D$-meson leading-twist DA $\phi_{2;D}$ should be updated. The criteria for determining the Borel windows of $\left<\xi^{n=1,\cdots,4}\right>_D$ is exhibited in Table \ref{t2criterion}, the Borel windows and the allowable regions for $\left<\xi^{n=1,\cdots,4}\right>_D$ are displayed in Table \ref{t2bw}. Then the values of those moments are updated as
\begin{eqnarray}
\left<\xi^1\right>_D|_{2 {\rm GeV}} &=& -0.421^{+0.025}_{-0.026}, \nonumber\\
\left<\xi^2\right>_D|_{2 {\rm GeV}} &=& 0.316^{+0.023}_{-0.021}, \nonumber\\
\left<\xi^3\right>_D|_{2 {\rm GeV}} &=& -0.186^{+0.015}_{-0.015}, \nonumber\\
\left<\xi^4\right>_D|_{2 {\rm GeV}} &=& 0.153^{+0.011}_{-0.010}
\label{xin}
\end{eqnarray}

\begin{table*}[htb]
\caption{Typical values for the model parameters of the $D$-meson leading-twist DAs at the scale $\mu =2{\rm GeV}$.}
\begin{tabular}{ c c c c | c c c c c c}
\hline
~$\left<\xi^1\right>_D$~& ~$\left<\xi^2\right>_D$~ & ~$\left<\xi^3\right>_D$~ & ~$\left<\xi^4\right>_D$~& ~$A_D({\rm GeV}^{-1})$~ & ~$B^{D}_1$~ & ~$B^{D}_2$~& ~$B^{D}_3$~ & ~$B^{D}_4$~ & ~$\beta_D({\rm GeV})$~ \\
\hline
~$-0.421$~& ~$0.316$~ & ~$-0.186$~ & ~$0.153$~& ~$16.071$~ & ~$-0.561$~ & ~$0.356$~& ~$-0.012$~ & ~$-0.093$~ & ~$0.986$~ \\
~$-0.421^{+0.025}$~& ~$0.316_{-0.021}$~ & ~$-0.186^{+0.015}$~ & ~$0.153_{-0.010}$~& ~$27.261$~ & ~$-0.287$~ & ~$0.418$~& ~$0.112$~ & ~$0.031$~ & ~$0.842$~ \\
~$-0.421_{-0.026}$~& ~$0.316^{+0.023}$~ & ~$-0.186_{-0.015}$~ & ~$0.153^{+0.011}$~& ~$8.360$~ & ~$-0.748$~ & ~$0.345$~& ~$-0.052$~ & ~$-0.201$~ & ~$1.350$~ \\
\hline
\end{tabular}
\label{T2DAparameter}
\end{table*}

\begin{figure}[htb]
\centering
\includegraphics[width=0.45\textwidth]{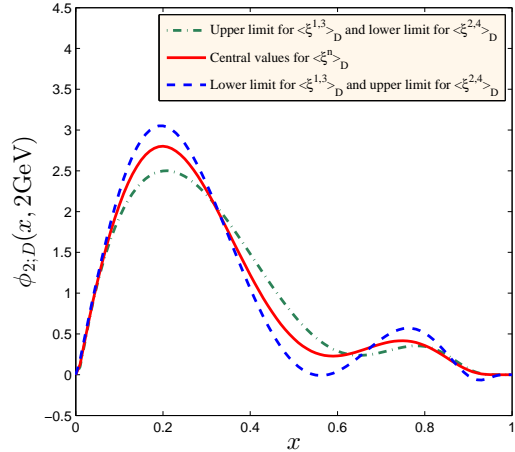}
\caption{The $D$-meson leading-twist DA $\phi_{2;D}$ corresponds to the input parameter values listed in Table \ref{T2DAparameter}. }
\label{fDAan}
\end{figure}

With the values of $\left<\xi^{n=1,\cdots,4}\right>_D$ shown in Eq.(\ref{xin}), the input parameters of the model (\ref{phi}) for the $D$-meson leading-twist DA $\phi_{2;D}$ can be obtained, and their typical values at the scale $\mu = 2{\rm GeV}$ are shown in Table \ref{T2DAparameter}. The corresponding curves of $\phi_{2;D}$ are shown in Fig.\ref{fDAan}. Comparing with the old simplified model suggested in Ref.\cite{Zhang:2017rwz}, the improved model (\ref{phi}) has a more obvious double-humped behavior and is narrower, both of which have a the peak around $x \sim 0.2$. Substituting the model parameters exhibited in Table \ref{T2DAparameter} into Eq.(\ref{K2}), one can obtain $\left<\bf{k}_\perp^2\right>_D^{1/2} \simeq (651 - 1038) {\rm MeV}$ (the central value is $755{\rm MeV}$). The behaviors of the twist-3 DAs is insensitive to the average value of the squared transverse momentum~\cite{Zhong:2014jla}, we will take the central values of $\left<\xi^{n=1,\cdots,4}\right>_D$, corresponding to $\left<\bf{k}_\perp^2\right>_D^{1/2} = 755 {\rm MeV}$, to constrain the behaviors of $\phi_{3;D}^p$ and $\phi_{3;D}^\sigma$ in later subsections.

\subsection{Moments of the $D$-meson twist-3 DAs}

\begin{table}[htb]
\caption{Criteria for determining the Borel windows of $\mu_D^{p}$ and $\mu_D^{\sigma}$, $\left<\xi^{n=1,\cdots,4}_p\right>_D$ and $\left<\xi^{n=1,\cdots,4}_\sigma\right>_D$. }
\begin{tabular}{ c  c  c }
\hline
~~ & ~Continue~ & ~Dimension-six~ \\
~~ & ~Contribution ($\%$)~ & ~Contribution ($\%$)~ \\
\hline
~$\mu_D^p$~ & ~$<30$~ & ~$<2$~ \\
~$\mu_D^\sigma$~ & ~$<30$~ & ~$<10$~ \\
\hline
~$\left<\xi^1_p\right>_D$~ & ~$<15$~ & ~$<5$~ \\
~$\left<\xi^2_p\right>_D$~ & ~$<30$~ & ~$<10$~ \\
~$\left<\xi^3_p\right>_D$~ & ~$<30$~ & ~$<10$~ \\
~$\left<\xi^4_p\right>_D$~ & ~$<45$~ & ~$<15$~ \\
\hline
~$\left<\xi^1_\sigma\right>_D$~ & ~$<30$~ & ~$<10$~ \\
~$\left<\xi^2_\sigma\right>_D$~ & ~$<30$~ & ~$<10$~ \\
~$\left<\xi^3_\sigma\right>_D$~ & ~$<45$~ & ~$<15$~ \\
~$\left<\xi^4_\sigma\right>_D$~ & ~$<45$~ & ~$<15$~ \\
\hline
\end{tabular}
\label{tcriterion}
\end{table}

\begin{table}[htb]
\caption{The Borel windows and the allowable regions for $\mu_D^{p}$, $\mu_D^{\sigma}$, $\left<\xi^n_p\right>_D$ and $\left<\xi^n_\sigma\right>_D$. All other input parameters are set to be their central values. }
\begin{tabular}{ c  c  c }
\hline
~~ & ~$M^2$~ & ~Value~ \\
\hline
~$\mu_D^p$~ & ~$[1.102,1.979]$~ & ~$[2.028,2.298]$~ \\
~$\mu_D^\sigma$~ & ~$[1.139,1.643]$~ & ~$[2.086,2.005]$~ \\
\hline
~$\left<\xi^1_p\right>_D$~ & ~$[1.478,2.111]$~ & ~$[-0.581,-0.496]$~ \\
~$\left<\xi^2_p\right>_D$~ & ~$[1.684,2.295]$~ & ~$[0.431,0.389]$~ \\
~$\left<\xi^3_p\right>_D$~ & ~$[2.372,2.961]$~ & ~$[-0.299,-0.281]$~ \\
~$\left<\xi^4_p\right>_D$~ & ~$[2.380,3.203]$~ & ~$[0.249,0.240]$~ \\
\hline
~$\left<\xi^1_\sigma\right>_D$~ & ~$[1.576,2.466]$~ & ~$[-0.504,-0.387]$~ \\
~$\left<\xi^2_\sigma\right>_D$~ & ~$[1.995,2.141]$~ & ~$[0.321,0.304]$~ \\
~$\left<\xi^3_\sigma\right>_D$~ & ~$[2.083,3.572]$~ & ~$[-0.248,-0.168]$~ \\
~$\left<\xi^4_\sigma\right>_D$~ & ~$[2.533,3.127]$~ & ~$[0.176,0.147]$~ \\
\hline
\end{tabular}
\label{tbw}
\end{table}

As suggested by Refs.\cite{Huang:2004tp, Huang:2005av}, the quarks inside the bound-state are not exactly on shell, and a more reasonable prediction on $\mu_\pi^{p}$ or $\mu_\pi^{\sigma}$ could be achieved by using the sum rules derived from the $0_{\rm th}$ moment of the pion twist-3 DA. More explicitly, By taking $n=0$ in sum rules (\ref{srxips}) and (\ref{srxipt}) and using the normalization conditions $\left<\xi_p^0\right>_D = \left<\xi_\sigma^0\right>_D = 1$, we obtain the sum rules for $\mu_\pi^{p}$ or $\mu_\pi^{\sigma}$. We present the criteria for determining the Borel window in Table \ref{tcriterion}, where for convenience we have also presented the criteria for the moments $\left<\xi^{n=1,\cdots,4}_p\right>_D$ and $\left<\xi^{n=1,\cdots,4}_\sigma\right>_D$. The determined Borel windows together with the determined values of $\mu_D^p$ and $\mu_D^\sigma$ are presented in Table \ref{tbw}. Table \ref{tbw} shows
\begin{eqnarray}
\mu_D^p &=& 2.535^{+0.136}_{-0.131} {\rm GeV}, \label{muD1} \\
\mu_D^\sigma &=& 2.534^{+0.267}_{-0.246} {\rm GeV}, \label{muD2}
\end{eqnarray}
where the errors are squared average of those from the errors of the parameters such as the Borel parameter, the condensates and the bound-state parameters. As a comparison, if roughly using the equation of motion for the on-shell particles~\cite{Braun:1989iv, Ball:1998sk, Ball:1998je}, we obtain $\mu_D^{p}=\mu_D^{\sigma}\simeq m_D^2/m_c \sim 3.04$ GeV, which are about $20\%$ larger than the sum rules predictions.

\begin{figure}[htb]
\centering
\includegraphics[width=0.45\textwidth]{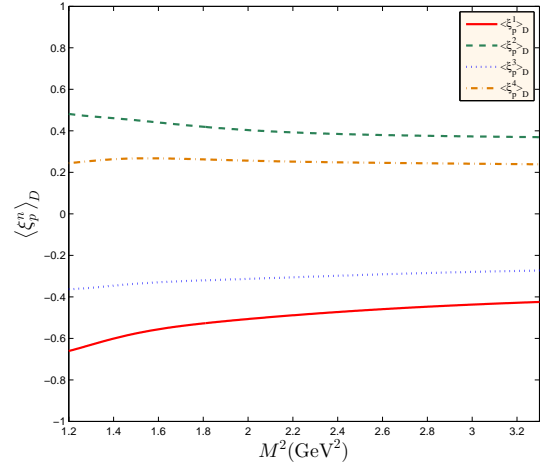}
\includegraphics[width=0.45\textwidth]{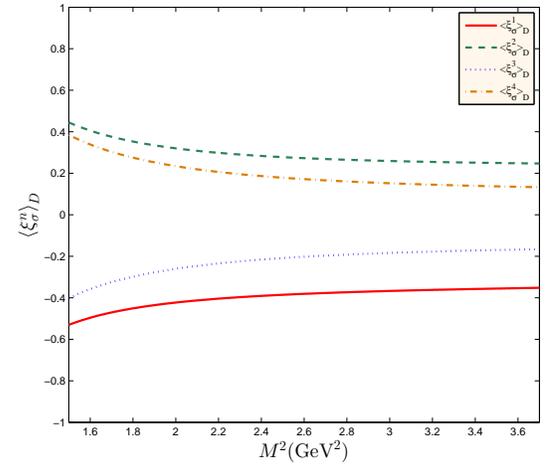}
\caption{The $D$-meson twist-3 DAs moments $\left<\xi_p^n\right>_D$ and $\left<\xi_\sigma^n\right>_D$ $(n=1,2,3,4)$ versus the Borel parameter $M^2$, where all input parameters are set to be their central values. The solid, the dashed, the dotted and the dash-dotted lines are for the first, the second, the third and the fourth moments, respectively. }
\label{fxin}
\end{figure}

We present the criteria for determining the Borel windows of the moments $\left<\xi_p^n\right>_D$ and $\left<\xi_\sigma^n\right>_D\ (n=1,2,3,4)$ in Table \ref{tcriterion}. The determined Borel windows and the corresponding moments are displayed in Table \ref{tbw}. Fig.\ref{fxin} shows the stabilities of the $D$-meson twist-3 DAs moments $\left<\xi_p^n\right>_D$ and $\left<\xi_\sigma^n\right>_D$ $(n=1,2,3,4)$ within the allowable Borel windows.

\begin{table*}[htb]
\caption{The impact of various inputs on the moments $\left<\xi_p^n\right>_D$ and $\left<\xi_\sigma^n\right>_D$. The Borel parameter $M$ is fixed to be its central value. The labels ``$|_{\rm up}$'' and ``$|_{\rm low}$'' stand for the upper and lower bounds of the inputs, and the symbols ``$+$'' and ``$-$'' represent the positive and negative errors brought by the corresponding input, respectively. The $\left<G^2\right>$, $\left<\bar{q}Gq\right>$ and $\left<G^3\right>$ are  abbreviations of the vacuum condensates $\left<\alpha_sG^2\right>$, $\left<g_s\bar{q}\sigma TGq\right>$ and $\left<g_s^3fG^3\right>$ respectively.}
\begin{tabular}{ l l l l l l l l l l l }
\hline
~ & ~$\left<G^2\right>|_{\rm up}$~ & ~$\left<G^2\right>|_{\rm low}$~ & ~$\left<G^3\right>|_{\rm up}$~ & ~$\left<G^3\right>|_{\rm low}$~ & ~$\left<\bar{q}q\right>|_{\rm up}$~ & ~$\left<\bar{q}q\right>|_{\rm low}$~ & ~$\left<\bar{q}Gq\right>|_{\rm up}$~ & ~$\left<\bar{q}Gq\right>|_{\rm low}$~ & ~$\mu_D^p|_{\rm up}$~ & ~$\mu_D^p|_{\rm low}$~ \\
\hline
~$\left<\xi^1_p\right>_D$~ & ~$-$~ & ~$+$~ & ~$/$~ & ~$/$~ & ~$+$~ & ~$-$~ & ~$-$~ & ~$+$~ & ~$+$~ & ~$-$~ \\
~$\left<\xi^2_p\right>_D$~ & ~$+$~ & ~$-$~ & ~$/$~ & ~$/$~ & ~$-$~ & ~$+$~ & ~$+$~ & ~$-$~ & ~$-$~ & ~$+$~ \\
~$\left<\xi^3_p\right>_D$~ & ~$-$~ & ~$+$~ & ~$/$~ & ~$/$~ & ~$+$~ & ~$-$~ & ~$-$~ & ~$+$~ & ~$+$~ & ~$-$~ \\
~$\left<\xi^4_p\right>_D$~ & ~$+$~ & ~$-$~ & ~$/$~ & ~$/$~ & ~$-$~ & ~$+$~ & ~$+$~ & ~$-$~ & ~$-$~ & ~$+$~ \\
~$\left<\xi^1_\sigma\right>_D$~ & ~$-$~ & ~$+$~ & ~$/$~ & ~$/$~ & ~$-$~ & ~$+$~ & ~$+$~ & ~$-$~ & ~$+$~ & ~$-$~ \\
~$\left<\xi^2_\sigma\right>_D$~ & ~$+$~ & ~$-$~ & ~$/$~ & ~$/$~ & ~$+$~ & ~$-$~ & ~$-$~ & ~$+$~ & ~$-$~ & ~$+$~ \\
~$\left<\xi^3_\sigma\right>_D$~ & ~$-$~ & ~$+$~ & ~$/$~ & ~$/$~ & ~$-$~ & ~$+$~ & ~$+$~ & ~$-$~ & ~$+$~ & ~$-$~ \\
~$\left<\xi^4_\sigma\right>_D$~ & ~$+$~ & ~$-$~ & ~$/$~ & ~$/$~ & ~$+$~ & ~$-$~ & ~$-$~ & ~$+$~ & ~$-$~ & ~$+$~ \\
\hline
~ & ~$\bar{m}_c|_{\rm up}$~ & ~$\bar{m}_c|_{\rm low}$~ & ~$\bar{m}_d|_{\rm up}$~ & ~$\bar{m}_d|_{\rm low}$~ & ~$m_D|_{\rm up}$~ & ~$m_D|_{\rm low}$~ & ~$f_D|_{\rm up}$~ & ~$f_D|_{\rm low}$~ & ~$\mu_D^\sigma|_{\rm up}$~ & ~$\mu_D^\sigma|_{\rm low}$~ \\
\hline
~$\left<\xi^1_p\right>_D$~ & ~$+$~ & ~$-$~ & ~$+$~ & ~$-$~ & ~$-$~ & ~$+$~ & ~$+$~ & ~$-$~ & ~$/$~ & ~$/$~ \\
~$\left<\xi^2_p\right>_D$~ & ~$-$~ & ~$+$~ & ~$-$~ & ~$+$~ & ~$+$~ & ~$-$~ & ~$-$~ & ~$+$~ & ~$/$~ & ~$/$~ \\
~$\left<\xi^3_p\right>_D$~ & ~$+$~ & ~$-$~ & ~$+$~ & ~$-$~ & ~$-$~ & ~$+$~ & ~$+$~ & ~$-$~ & ~$/$~ & ~$/$~ \\
~$\left<\xi^4_p\right>_D$~ & ~$-$~ & ~$+$~ & ~$-$~ & ~$+$~ & ~$+$~ & ~$-$~ & ~$-$~ & ~$+$~ & ~$/$~ & ~$/$~ \\
~$\left<\xi^1_\sigma\right>_D$~ & ~$+$~ & ~$-$~ & ~$+$~ & ~$-$~ & ~$-$~ & ~$+$~ & ~$+$~ & ~$-$~ & ~$+$~ & ~$-$~ \\
~$\left<\xi^2_\sigma\right>_D$~ & ~$-$~ & ~$+$~ & ~$-$~ & ~$+$~ & ~$+$~ & ~$-$~ & ~$-$~ & ~$+$~ & ~$-$~ & ~$+$~ \\
~$\left<\xi^3_\sigma\right>_D$~ & ~$+$~ & ~$-$~ & ~$+$~ & ~$-$~ & ~$-$~ & ~$+$~ & ~$+$~ & ~$-$~ & ~$+$~ & ~$-$~ \\
~$\left<\xi^4_\sigma\right>_D$~ & ~$-$~ & ~$+$~ & ~$-$~ & ~$+$~ & ~$+$~ & ~$-$~ & ~$-$~ & ~$+$~ & ~$-$~ & ~$+$~ \\
\hline
\end{tabular}
\label{tinputs}
\end{table*}

Following the same idea suggested by Ref.\cite{Zhang:2017rwz}, we analyze the impact of various inputs on the moments $\left<\xi_p^n\right>_D$ and $\left<\xi_\sigma^n\right>_D$, the results are put in Table~\ref{tinputs}. Table~\ref{tinputs} shows that the effects of the input parameters on $\left<\xi_p^n\right>_D$ and $\left<\xi_\sigma^n\right>_D$ are similar to those of the leading-twist moments $\left<\xi^n\right>_D$~\cite{Zhang:2017rwz}. By varying the mentioned error sources within allowable regions, we obtain
\begin{eqnarray}
\left<\xi_p^1\right>_D|_{2 {\rm GeV}} &=& -0.484^{+0.075}_{-0.080}, \nonumber\\
\left<\xi_p^2\right>_D|_{2 {\rm GeV}} &=& 0.400^{+0.057}_{-0.052}, \nonumber\\
\left<\xi_p^3\right>_D|_{2 {\rm GeV}} &=& -0.277^{+0.037}_{-0.041}, \nonumber\\
\left<\xi_p^4\right>_D|_{2 {\rm GeV}} &=& 0.242^{+0.035}_{-0.033}
\label{xinp}
\end{eqnarray}
and
\begin{eqnarray}
\left<\xi_\sigma^1\right>_D|_{2 {\rm GeV}} &=& -0.381^{+0.068}_{-0.071}, \nonumber\\
\left<\xi_\sigma^2\right>_D|_{2 {\rm GeV}} &=& 0.296^{+0.037}_{-0.033}, \nonumber\\
\left<\xi_\sigma^3\right>_D|_{2 {\rm GeV}} &=& -0.190^{+0.043}_{-0.044}, \nonumber\\
\left<\xi_\sigma^4\right>_D|_{2 {\rm GeV}} &=& 0.156^{+0.024}_{-0.022},
\label{xins}
\end{eqnarray}
where the errors are squared averages of the errors from all the mentioned error sources. The errors are dominated by the parameters $\mu_D^{p,\sigma}$, $f_D$, $m_c$, and the condensates $\left<\bar{q}q\right>$ and $\left<g_s\bar{q}\sigma TGq\right>$.

\subsection{Properties of the twist-3 DAs $\phi^p_{3;D}$ and $\phi^\sigma_{3;D}$}

\begin{table*}[htb]
\caption{Typical values for the input parameters of the $D$-meson twist-3 DAs at the scale $\mu =2{\rm GeV}$.}
\begin{tabular}{ c c c c | c c c c c c}
\hline
~$\left<\xi_p^1\right>_D$~& ~$\left<\xi_p^2\right>_D$~ & ~$\left<\xi_p^3\right>_D$~ & ~$\left<\xi_p^4\right>_D$~& ~$A_D^p({\rm GeV}^{-1})$~ & ~$B^{D,p}_1$~ & ~$B^{D,p}_2$~& ~$B^{D,p}_3$~ & ~$B^{D,p}_4$~ & ~$\beta_D^p({\rm GeV})$~ \\
\hline
~$-0.484$~& ~$0.400$~ & ~$-0.277$~ & ~$0.242$~& ~$34.907$~ & ~$-0.927$~ & ~$2.522$~& ~$-0.496$~ & ~$1.463$~ & ~$0.993$~ \\
~$-0.484^{+0.075}$~& ~$0.400_{-0.052}$~ & ~$-0.277^{+0.037}$~ & ~$0.242_{-0.033}$~& ~$21.744$~ & ~$-1.195$~ & ~$2.185$~& ~$-0.788$~ & ~$1.520$~ & ~$1.131$~ \\
~$-0.484_{-0.080}$~& ~$0.400^{+0.057}$~ & ~$-0.277_{-0.041}$~ & ~$0.242^{+0.035}$~& ~$91.860$~ & ~$-0.068$~ & ~$2.886$~& ~$0.255$~ & ~$1.511$~ & ~$0.804$~ \\
\hline
~$\left<\xi_\sigma^1\right>_D$~& ~$\left<\xi_\sigma^2\right>_D$~ & ~$\left<\xi_\sigma^3\right>_D$~ & ~$\left<\xi_\sigma^4\right>_D$~& ~$A_D^\sigma({\rm GeV}^{-1})$~ & ~$B^{D,\sigma}_1$~ & ~$B^{D,\sigma}_2$~& ~$B^{D,\sigma}_3$~ & ~$B^{D,\sigma}_4$~ & ~$\beta_D^\sigma({\rm GeV})$~ \\
\hline
~$-0.381$~& ~$0.296$~ & ~$-0.190$~ & ~$0.156$~& ~$11.378$~ & ~$-0.564$~ & ~$0.270$~& ~$-0.104$~ & ~$0.064$~ & ~$1.148$~ \\
~$-0.381^{+0.068}$~& ~$0.296_{-0.033}$~ & ~$-0.190^{+0.043}$~ & ~$0.156_{-0.022}$~& ~$10.871$~ & ~$-0.400$~ & ~$0.216$~& ~$0.011$~ & ~$0.089$~ & ~$1.201$~ \\
~$-0.381_{-0.071}$~& ~$0.296^{+0.037}$~ & ~$-0.190_{-0.044}$~ & ~$0.156^{+0.024}$~& ~$12.860$~ & ~$-0.783$~ & ~$0.322$~& ~$-0.262$~ & ~$0.026$~ & ~$1.050$~ \\
\hline
\end{tabular}
\label{DAparameter}
\end{table*}

\begin{figure}[htb]
\centering
\includegraphics[width=0.45\textwidth]{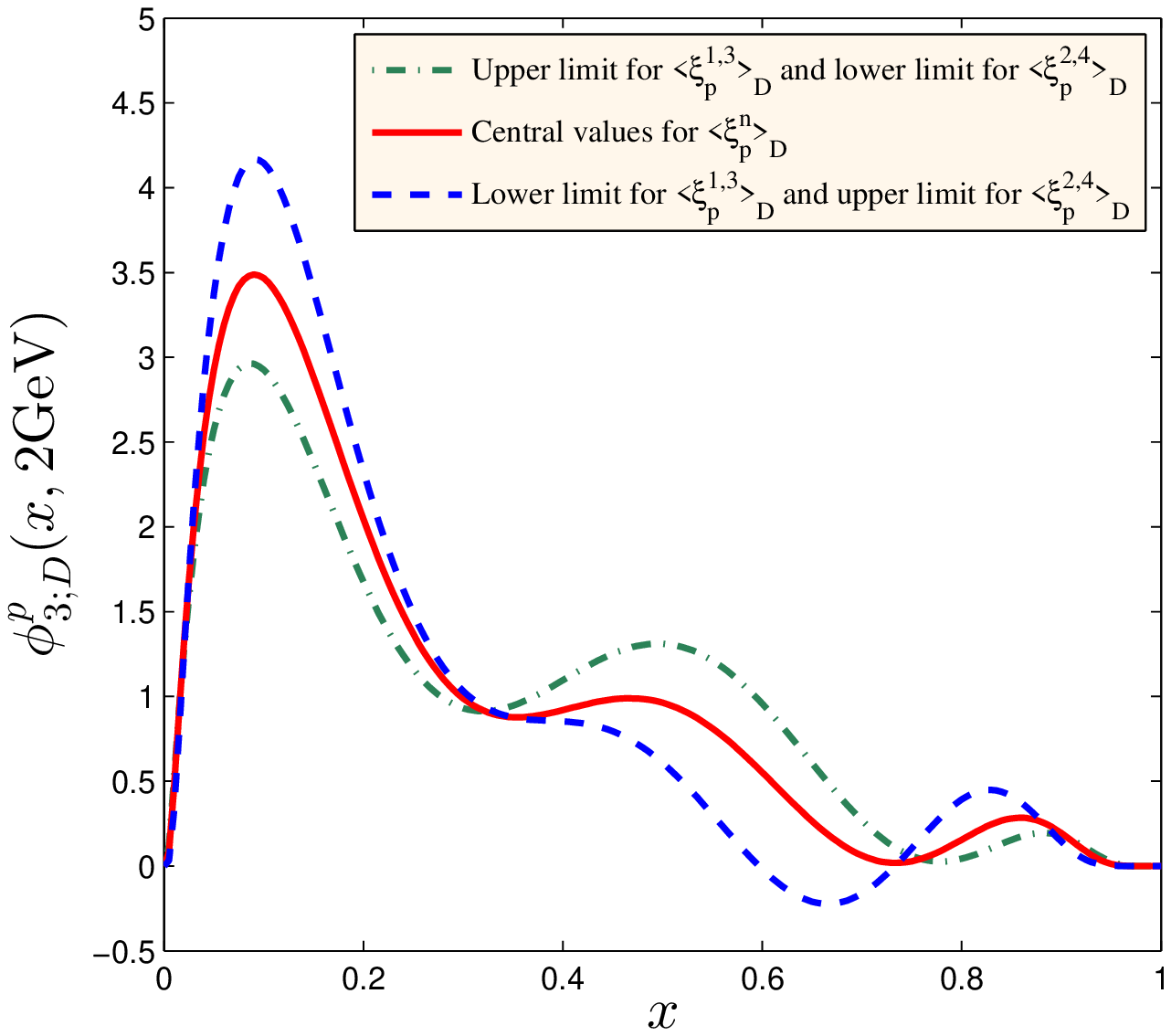}
\includegraphics[width=0.45\textwidth]{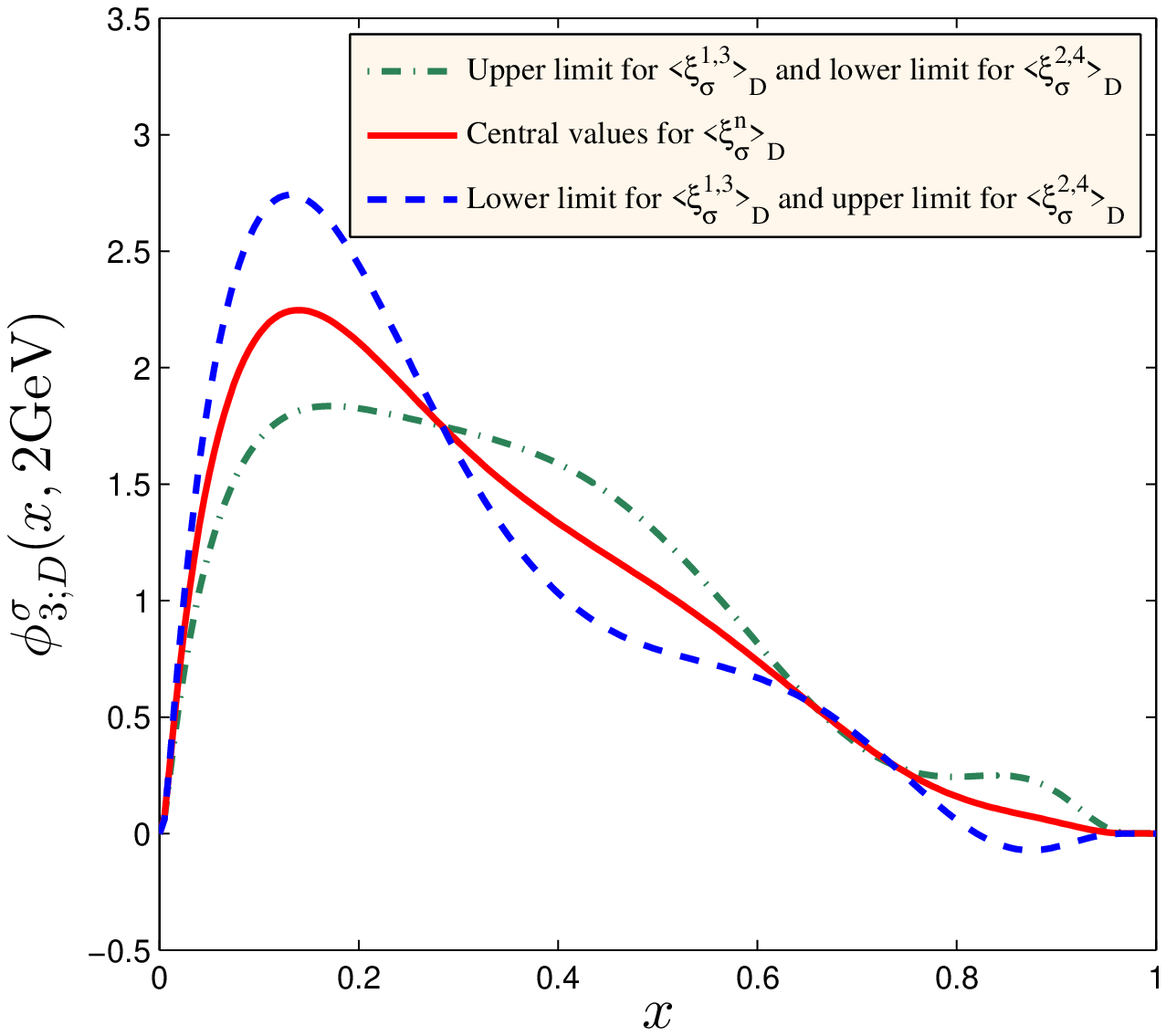}
\caption{The $D$-meson twist-3 DAs $\phi_{3;D}^p$ and $\phi_{3;D}^\sigma$ with the parameter values given in Table \ref{DAparameter}.}
\label{fD_T3DAs_xin}
\end{figure}

Similar to the leading-twist DA~\cite{Zhang:2017rwz}, the twist-3 DA moments cannot be varied independently. For example, if $\left<\xi_p^1\right>_D$ and $\left<\xi_p^3\right>_D$ take the upper bound, $\left<\xi_p^2\right>_D$ and $\left<\xi_p^4\right>_D$ should take the lower bound so as to achieve a self-consistent estimation of $\phi_{3;D}^p$ uncertainty. The error band of $\phi_{3;D}^p$ can be determined by two sets of $\left<\xi_p^n\right>_D$, namely, (i) $\left<\xi_p^1\right>_D|_{2{\rm GeV}} = -0.484^{+0.075}$, $\left<\xi_p^2\right>_D|_{2{\rm GeV}} = 0.400_{-0.052}$, $\left<\xi_p^3\right>_D|_{2{\rm GeV}} = -0.277^{+0.037}$, $\left<\xi_p^4\right>_D|_{2{\rm GeV}} = 0.242_{-0.033}$; (ii) $\left<\xi_p^1\right>_D|_{2{\rm GeV}} = -0.484_{-0.080}$, $\left<\xi_p^2\right>_D|_{2{\rm GeV}} = 0.400^{+0.057}$, $\left<\xi_p^3\right>_D|_{2{\rm GeV}} = -0.277_{-0.041}$, $\left<\xi_p^4\right>_D|_{2{\rm GeV}} = 0.242^{+0.035}$. The twist-3 DA $\phi_{3;D}^\sigma$ can be treated via the way. We present the determined values for the parameters of the twist-3 DAs $\phi_{3;D}^p$ and $\phi_{3;D}^\sigma$ at the scale $\mu=2{\rm GeV}$ in Table \ref{DAparameter}, and the corresponding cures are displayed in Fig.\ref{fD_T3DAs_xin}.

Comparing Eqs.(\ref{xinp}) and (\ref{xin}), one can find that the differences among the moments of $\phi_{3;D}^p$ and $\phi_{2;D}$ are about $13-37\%$, and which increases with the increase of the moment order, Figs.(\ref{fDAan}, \ref{fD_T3DAs_xin}) show that there is large difference between the behaviors of $\phi_{2;D}$ and $\phi_{3;D}^p$. It is then reasonable to assume that large discrepancy on the predictions involving them could be achieved by taking the rough approximation, $\phi_{3;D}^p \simeq \phi_{2;D}$.

\begin{figure}[htb]
\centering
\includegraphics[width=0.45\textwidth]{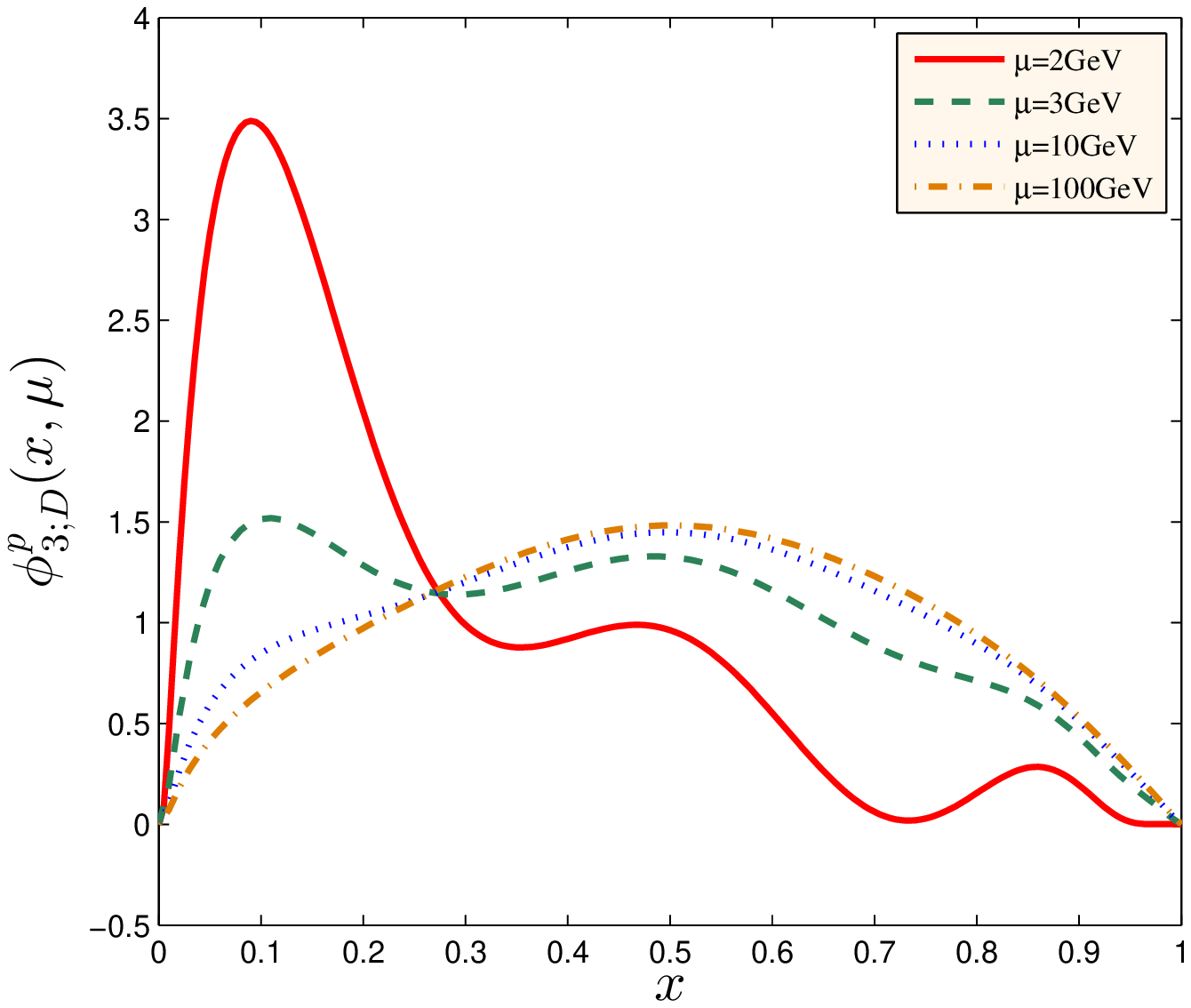}
\includegraphics[width=0.45\textwidth]{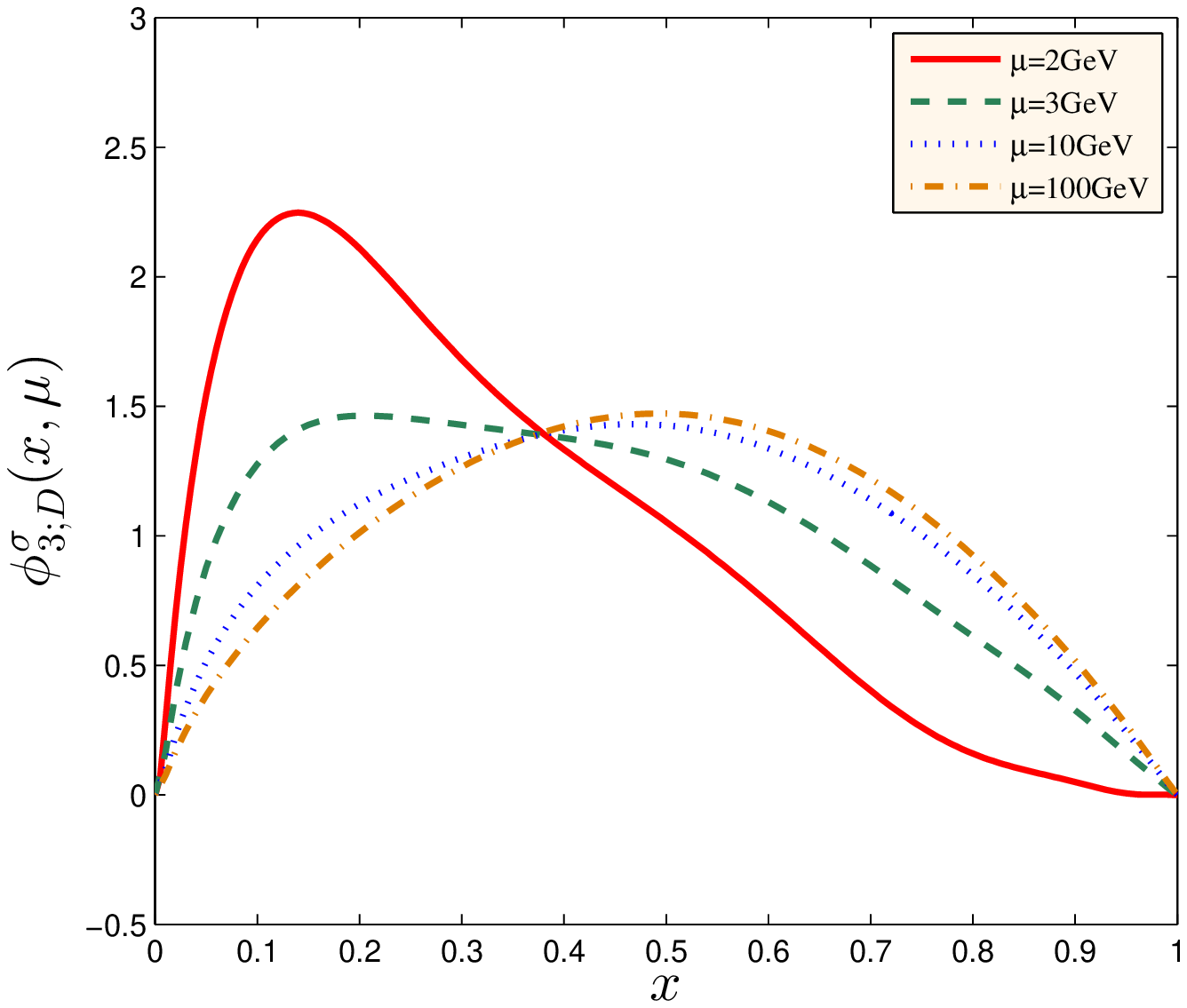}
\caption{The $D$-meson twist-3 DAs $\phi_{3;D}^p$ and $\phi_{3;D}^\sigma$ at different scales, where the solid, the dashed, the dotted and the dash-dotted lines are for the scales $\mu = 2$, $3$, $10$, $100$ GeV, respectively.}
\label{fD_T3DAs_Evolution}
\end{figure}

The $D$-meson twist-3 DAs $\phi_{3;D}^p$ and $\phi_{3;D}^\sigma$ at any other scales can be achieved by using the evolution equation. We present the twist-3 DAs $\phi_{3;D}^p$ and $\phi_{3;D}^\sigma$ under several typical scales, such as $\mu = 2$ ,$3$, $10$ and $100$ GeV in Fig.\ref{fD_T3DAs_Evolution}. With the increment of $\mu$, $\phi_{3;D}^p$ and $\phi_{3;D}^\sigma$ become more symmetric with the peak around $x=0.5$, both of which tend to the asymptotic form $6x(1-x)$. This situation in $\phi_{3;D}^p$ is different from that of the heavy pseudo-scalar meson, which shows a humped behavior near the end-point region $x\to 0,1$ for high scales~\cite{Zhong:2016kuv}. Unlike the asymptotic form of $\phi_{3;D}^p$, i.e. $\phi_{3;D}^p(x,\mu\to\infty) \equiv 1$, the model (\ref{phipmodel}) for $\mu\to\infty$ still equals to zero as $x\to 0$ or $x\to1$, which is due to exponential suppression from the BHL-prescription. It has already been observed that a more reasonable twist-3 contributions to the pion form factor~\cite{Huang:2004su} and the $B\to\pi$ TFFs~\cite{Huang:2004hw} can be achieved by using the pion twist-3 DAs with similar end-point behaviors. Thus the $D$-meson DAs with suitable end-point singularity behavior shall be helpful for achieving a more reliable twist-3 predictions within the pQCD approach.

\subsection{The $B\to D$ TFFs}

Substituting the $D$-meson twist-2 and twit-3 DAs into the LCSRs (\ref{srf+}, \ref{srf+-}), we can obtain the $B\to D$ TFFs $f^{B\to D}_{+,0}(q^2)$ with the help of Eq.(\ref{f0}). To do the numerical calculation, we take the $B$-meson mass $m_{\overline{B}^0} = 5279.63 \pm {\rm MeV}$, the decay constant $f_B = 188 \pm 17 \pm 18 {\rm MeV}$, and the $b$-quark mass $\bar{m}_b(\bar{m}_b) = 4.18^{+0.04}_{-0.03} {\rm GeV}$~\cite{PDG_Olive:2016xmw}. For the continuous threshold parameter $s_0^B$, we take it to be $s_0^B = 36 \pm 1 {\rm GeV^2}$; we take the Borel parameter $M^2 = (20 \sim 30) {\rm GeV^2}$, the factorization scale $\mu \simeq 3 {\rm GeV}$. We need to run the model parameters of the $D$-meson twist-2, 3 DAs exhibited in Table \ref{T2DAparameter} and \ref{DAparameter} up to the scale $\mu = 3\rm GeV$ via the QCD evolution equation, which are presented in Table \ref{DAparameter3}. It is found that the differences caused by different bound-state masses are less than $10^{-4}$ of the total contributions, thus TFFs $f^{B\to D}_{+,0}(q^2)$ obtained with the above parameters can be applied for both the decays $B^- \to D^0 l\bar{\nu}_{l}$ and $\overline{B}^0 \to D^+ l \bar{\nu}_l$.

\begin{table}[htb]
\caption{Typical values for the input parameters of the $D$-meson twist-2,3 DAs at the scale $\mu =3{\rm GeV}$.}
\begin{tabular}{ c c c c c c}
\hline
~$A_D({\rm GeV}^{-1})$~ & ~$B^{D}_1$~ & ~$B^{D}_2$~& ~$B^{D}_3$~ & ~$B^{D}_4$~ & ~$\beta_D({\rm GeV})$~ \\
~$8.293$~ & ~$-0.349$~ & ~$0.231$~& ~$-0.007$~ & ~$-0.052$~ & ~$1.309$~ \\
~$6.724$~ & ~$-0.364$~ & ~$0.202$~& ~$0.018$~ & ~$-0.024$~ & ~$1.465$~ \\
~$9.231$~ & ~$-0.313$~ & ~$0.258$~& ~$-0.009$~ & ~$-0.071$~ & ~$1.249$~ \\
\hline
~$A_D^p({\rm GeV}^{-1})$~ & ~$B^{D,p}_1$~ & ~$B^{D,p}_2$~& ~$B^{D,p}_3$~ & ~$B^{D,p}_4$~ & ~$\beta_D^p({\rm GeV})$~ \\
\hline
~$10.387$~ & ~$-0.491$~ & ~$1.408$~& ~$-0.246$~ & ~$0.771$~ & ~$1.401$~ \\
~$7.451$~ & ~$-0.596$~ & ~$1.136$~& ~$-0.324$~ & ~$0.762$~ & ~$1.602$~ \\
~$23.402$~ & ~$-0.009$~ & ~$1.878$~& ~$0.105$~ & ~$0.989$~ & ~$1.078$~ \\
\hline
~$A_D^\sigma({\rm GeV}^{-1})$~ & ~$B^{D,\sigma}_1$~ & ~$B^{D,\sigma}_2$~& ~$B^{D,\sigma}_3$~ & ~$B^{D,\sigma}_4$~ & ~$\beta_D^\sigma({\rm GeV})$~ \\
\hline
~$4.431$~ & ~$-0.316$~ & ~$0.122$~& ~$-0.037$~ & ~$0.029$~ & ~$1.907$~ \\
~$5.018$~ & ~$-0.266$~ & ~$0.108$~& ~$0.005$~ & ~$0.040$~ & ~$1.749$~ \\
~$6.007$~ & ~$-0.320$~ & ~$0.188$~& ~$-0.077$~ & ~$0.030$~ & ~$1.551$~ \\
\hline
\end{tabular}
\label{DAparameter3}
\end{table}

At the maximum recoil point $q^2 = 0$, we have:
\begin{eqnarray}
f^{B\to D}_{+,0}(0) = 0.570\ ^{+0.029}_{-0.032}|_{\rm LO} + 0.089|_{\rm NLO},
\label{tff0}
\end{eqnarray}
where the error is obtained by adding up of all the errors in quadrature, whose error sources contain the choices of $\phi_{2;D}$, $\phi_{3;D}^p$ and $\phi_{3;D}^\sigma$, the Borel parameter $M^2$, the continuum threshold $s_0^B$, the $B(D)$-meson decay constant $f_{B(D)}$, the $b$-quark mass $m_b$ and the normalization parameter $\mu_D^{p(\sigma)}$. For the LO contributions, we have found that different choices of the $D$-meson DAs $\phi_{2;D}$, $\phi_{3;D}^p$ and $\phi_{3;D}^\sigma$ shall bring about $(0.1\sim 0.5)\%$, $(1.5\sim 2.1)\%$ and $(0.1\sim 0.2)\%$ errors to $f^{B\to D}_{+,0}(0)$, respectively. Thus more precise twist-2 DA $\phi_{2;D}$ and the twist-3 DA $\phi_{3;D}^p$ are important for a precise prediction on the $B\to D$ TFFs.

\begin{figure}[htb]
\centering
\includegraphics[width=0.45\textwidth]{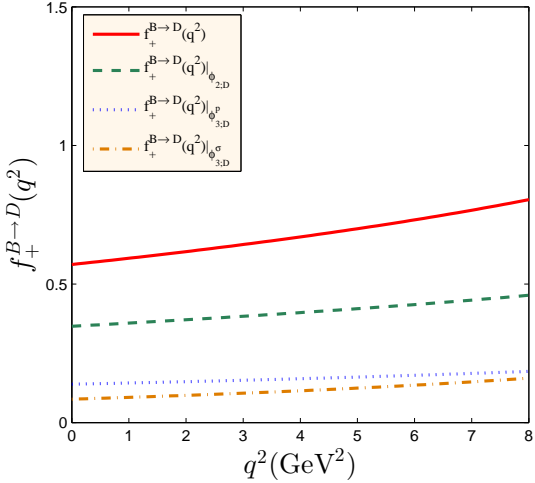}
\caption{The LO LCSR prediction on the TFF $f^{B\to D}_{+}(q^2)$, where the solid line is for the total LO TFF $f^{B\to D}_{+}(q^2)$; the dashed, the dash-dot and the dotted lines are for the separate contributions from $\phi_{2;D}$, $\phi_{3;D}^p$ and $\phi_{3;D}^\sigma$, respectively.}
\label{ftff1q_parts}
\end{figure}

To show how various $D$-meson DAs contribute to the TFF, we present the LO contributions to the TFF $f^{B\to D}_{+}(q^2)$ separately from $\phi_{2;D}$, $\phi_{3;D}^p$ and $\phi_{3;D}^\sigma$ in Fig.\ref{ftff1q_parts}, in which all input parameters are set to be their central values. In whole $q^2$-region, the twist-3 contributions are sizable but smaller than the twist-2 contribution but sizable. For example, at $q^2=0$, we have $f^{B\to D}_{+}(0)|_{\phi_{2;D}} = 0.347$, $f^{B\to D}_{+}(0)|_{\phi_{3;D}^p} = 0.138$ and $f^{B\to D}_{+}(0)|_{\phi_{3;D}^\sigma} = 0.085$, which provide $61\%$, $24\%$ and $15\%$ contribution to the LO TFF $f^{B\to D}_{+}(0)$, respectively.

\begin{figure}[htb]
\centering
\includegraphics[width=0.45\textwidth]{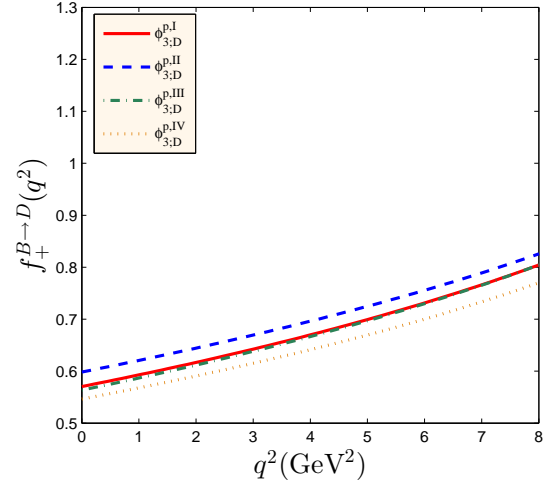}
\caption{The LCSR prediction on the TFF $f^{B\to D}_{+}(q^2)$ by using four different twist-3 DA $\phi_{3;D}^p$, where the solid, the dashed, the dash-dot, and the dotted lines are for $\phi_{3;D}^{p, I}$, $\phi_{3;D}^{p, II}$, $\phi_{3;D}^{p, III}$ and $\phi_{3;D}^{p, IV}$, respectively.}
\label{ftff1q_comparison}
\end{figure}

To show how various twist-3 DA $\phi_{3;D}^p$ models affect the LO TFF, we take four models for the twist-3 DA $\phi_{3;D}^p$, e.g. I) $\phi_{3;D}^{p, I}$ which equals to our present model (\ref{phipmodel}); II) $\phi_{3;D}^{p, II}= \phi_{2;D}$ with $\phi_{2;D}$ from Eq.(\ref{phi}); III) $\phi_{3;D}^{p, III}= \phi_{2;D}$ with $\phi_{2;D}$ equals to the KLS model~\cite{MODELI_Keum:2003js}; IV) $\phi_{3;D}^{p, IV}\equiv 1$. We present such a comparison in Fig.\ref{ftff1q_comparison}, in which all other parameters are set to be their central values. Fig.\ref{ftff1q_comparison} shows that the TFF is sensitive to the behavior of $\phi_{3;D}^p$. For example, at the large recoil point, by taking $\phi_{3;D}^p=\phi_{3;D}^{p, II}$, we obtain $f^{B\to D}_{+}(0) = 0.598$, which is $4.9\%$ larger than the value derived by taking $\phi_{3;D}^p=\phi_{3;D}^{p, I}$; by taking $\phi_{3;D}^p=\phi_{3;D}^{p, IV}$, we obtain $f^{B\to D}_{+}(0) = 0.546$, which is $4.2\%$ smaller than the value derived by taking $\phi_{3;D}^p=\phi_{3;D}^{p, I}$. Moreover, by taking $\phi_{3;D}^p=\phi_{3;D}^{p, III}$, we obtain $f^{B\to D}_{+}(0) = 0.563$, which is close to the value derived by taking $\phi_{3;D}^p=\phi_{3;D}^{p, I}$, the reason is that the behavior of $\phi_{3;D}^{p, I}$ at $\mu = 3\rm GeV$ is coincidentally close to $\phi_{3;D}^{p, III}$.

\subsection{The ratio $\mathcal{R}(D)$}

\begin{table}[htb]
\caption{The fitted parameters $a_{+(0)}$ and $b_{+(0)}$ for the extrapolation of the TFFs $f^{B\to D}_{+(0)}(q^2)$. }
\begin{tabular}{ c | c | c | c | c }
\hline
~$f^{B\to D}_{+,0}(0)$~ & ~$a_+$~ & ~$b_+$~ & ~$a_0$~ & ~$b_0$~ \\
\hline
~$0.689$~ & ~$1.036133$~ & ~$-0.057093$~ & ~$0.108209$~ & ~$-1.362107$~ \\
~$0.659$~ & ~$1.040720$~ & ~$-0.067793$~ & ~$0.100657$~ & ~$-1.381510$~ \\
~$0.627$~ & ~$1.039425$~ & ~$-0.109856$~ & ~$0.082498$~ & ~$-1.447109$~ \\
\hline
\end{tabular}
\label{tfwr}
\end{table}

The LCSRs for the TFFs $f^{B\to D}_{+,0}(q^2)$ are reliable in low and intermediate regions such as $q^2\in[0,8]{\rm GeV}^2$, and to make it applicable in all $q^2$-region, one usually extrapolates it by using the following parametrization~\cite{Wang:2008xt}
\begin{eqnarray}
f^{B\to D}_{+(0)}(q^2) = \frac{f^{B\to D}_{+(0)}(0)}{1 - a_{+(0)}\left(q^2/m_B^2\right) + b_{+(0)}\left(q^2/m_B^2\right)^2}.
\label{fwr}
\end{eqnarray}
On the other hand, the LQCD results for the TFFs $f^{B\to D}_{+,0}(q^2)$ are available for high energy region~\cite{LQCD_Lattice:2015rga, LQCD_Na:2015kha}, thus one may combine the LCSR and LQCD predictions to accurate a reliable prediction within the whole $q^2$-region. In doing the combination, we adopt the extrapolation formulae (\ref{fwr}) to fit our LCSR predictions for the TFFs with $\phi_{3;D}^{p, I}$ and the LQCD predictions by the HPQCD Collaboration~\cite{LQCD_Na:2015kha}. The fitted parameters $a_{+(0)}$ and $b_{+(0)}$ are presented in Table \ref{tfwr}.

\begin{figure}[htb]
\centering
\includegraphics[width=0.45\textwidth]{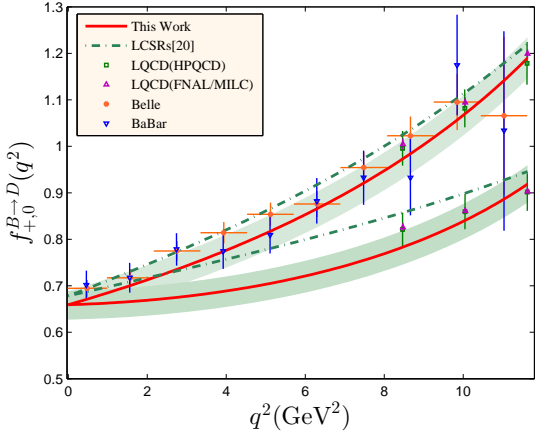}
\caption{The fitted LCSR and LQCD predictions on the TFFs $f^{B\to D}_{+,0}(q^2)$. The solid lines are central values of the TFFs $f^{B\to D}_{+,0}(q^2)$, and the shaded hands are their corresponding uncertainties. The extrapolated LCSR predictions with the vacuum-to-$B$-meson correlator~\cite{Wang:2017jow}, and the LQCD predictions by the HPQCD Collaboration~\cite{LQCD_Na:2015kha} or by the FNAL/MILC Collaboration~\cite{LQCD_Lattice:2015rga}, and the data from the Belle Collaboration~\cite{Glattauer:2015teq} and the BaBar Collaboration~\cite{Aubert:2008yv} are presented as a comparison. }
\label{ftffq_comparison}
\end{figure}

We present the fitting TFFs $f^{B\to D}_{+,0}(q^2)$ and their uncertainties in Fig.\ref{ftffq_comparison}. The solid lines are the central values of the TFFs $f^{B\to D}_{+}(q^2)$ and $f^{B\to D}_{0}(q^2)$ and the shaded hands are their uncertainties. The extrapolated LCSR predictions with the vacuum-to-$B$-meson correlator~\cite{Wang:2017jow}, and the LQCD predictions by the HPQCD Collaboration~\cite{LQCD_Na:2015kha} or by the FNAL/MILC Collaboration~\cite{LQCD_Lattice:2015rga}, and the data from the Belle and BaBar Collaborations~\cite{Glattauer:2015teq, Aubert:2008yv} are presented as a comparison. Fig.\ref{ftffq_comparison} shows our predictions on the TFFs $f^{B\to D}_{+}(q^2)$ agree with the Belle and BaBar measurements within errors.

\begin{figure}[htb]
\centering
\includegraphics[width=0.45\textwidth]{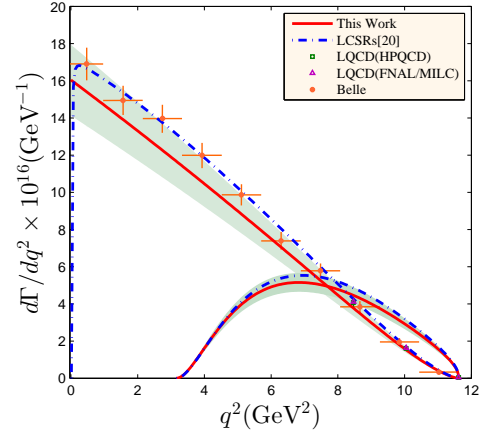}
\caption{Differential decay rates for the decay $\overline{B}^0 \to D^+l\bar{\nu}_{l}$. The solid lines are for $\overline{B}^0 \to D^+l^\prime\bar{\nu}_{l^\prime}$ and $\overline{B^0} \to D^+\tau\bar{\nu}_{\tau}$, respectively. The shaded hands are their uncertainties. The extrapolated LCSR prediction with the vacuum-to-$B$-meson correlation~\cite{Wang:2017jow}, the LQCD prediction by the HPQCD Collaboration~\cite{LQCD_Na:2015kha} and the FNAL/MILC Collaboration~\cite{LQCD_Lattice:2015rga} are presented as a comparison. The experimental data are from Belle Collaboration~\cite{Glattauer:2015teq}. }
\label{fdgamma}
\end{figure}

As a step forward, we present the differential decay rates for the decay $\overline{B}^0 \to D^+l\bar{\nu}_{l}$ in Fig.\ref{fdgamma}, where the solid lines are for $\overline{B}^0 \to D^+l^\prime\bar{\nu}_{l^\prime}$ and $\overline{B}^0 \to D^+\tau\bar{\nu}_{\tau}$, respectively. The shaded bands are their uncertainties. The extrapolated LCSR prediction with the vacuum-to-$B$-meson correlation~\cite{Wang:2017jow}, the LQCD prediction by the HPQCD Collaboration~\cite{LQCD_Na:2015kha} and the FNAL/MILC Collaboration~\cite{LQCD_Lattice:2015rga} are presented as a comparison.

\begin{table*}[htb]
\caption{Theoretical predictions for the branching ratios (in units of $10^{-2}$) of the $B\to D l\bar{\nu}_l$ decays. As a comparison, the PDG values~\cite{PDG_Olive:2016xmw}, the BaBar predictions~\cite{BABAR_Lees:2012xj, BABAR_Lees:2013uzd, Aubert:2009ac}, the HQET predictions~\cite{HQET_Fajfer:2012vx} are also presented.}
\begin{tabular}{ c c c c c }
\hline
~Channels~ & ~This work~ & ~HQET~ & ~BaBar~ & ~PDG~ \\
\hline
~$\overline{B}^0 \to D^+l^\prime\bar{\nu}_{l^\prime}$~ & ~$2.086^{+0.230}_{-0.232}$~ & ~$-$~ & ~$2.23\pm 0.16$~ & ~$2.19\pm 0.12$~ \\
~$\overline{B}^0 \to D^+\tau\bar{\nu}_\tau$~ & ~$0.666^{+0.058}_{-0.057}$~ & ~$0.64\pm 0.05$~ & ~$1.01\pm 0.22$~ & ~$1.03\pm 0.22$~  \\
~$B^- \to D^0 l^\prime\bar{\nu}_{l^\prime}$~ & ~$2.260^{+0.249}_{-0.251}$~ & ~$-$~ & ~$2.31\pm 0.12$~ & ~$2.27\pm 0.11$~ \\
~$B^- \to D^0 \tau\bar{\nu}_\tau$~ & ~$0.724^{+0.063}_{-0.062}$~ & ~$0.66\pm 0.05$~ & ~$0.99\pm 0.23$~ & ~$0.77\pm 0.25$~ \\
\hline
\end{tabular}
\label{tbr}
\end{table*}

\begin{figure}[htb]
\centering
\includegraphics[width=0.45\textwidth]{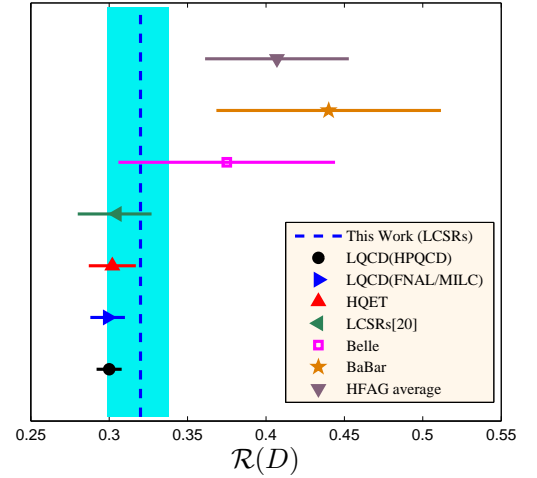}
\caption{The ratio $\mathcal{R}(D)$ of the semi-leptonic decays $B\to D l\bar{\nu}_l$. The dashed line stands for the central value and  the shaded band is its uncertainty.}
\label{fRD}
\end{figure}

We present the branching ratios for the decay $B\to D l\bar{\nu}_l$ in Table \ref{tbr}, where the PDG values~\cite{PDG_Olive:2016xmw}, the BaBar data~\cite{BABAR_Lees:2012xj, BABAR_Lees:2013uzd, Aubert:2009ac}, the HQET predictions~\cite{HQET_Fajfer:2012vx} are presented as a comparison. To do the numerical calculation, we adopt $G_F = 1.1663787(6)\times 10^{-5} {\rm GeV}^{-2}$, $|V_{cb}| = (40.5 \pm 1.5) \times 10^{-3}$, $m_\tau = 1776.86 \pm 0.12 {\rm MeV}$, $m_{B^-} = 5279.32 \pm 0.14 {\rm MeV}$, $m_{D^0} = 1864.83 \pm 0.05 {\rm MeV}$, $\tau_{\overline{B}^0} = (1.520 \pm 0.004) \times 10^{-12}s$ and $\tau_{B^-} = (1.638 \pm 0.004) \times 10^{-12} s$\cite{PDG_Olive:2016xmw}. Table~\ref{tbr} shows our predictions on the branching ratios $\mathcal{B}(\overline{B}^0 \to D^+ l^\prime\bar{\nu}_{l^\prime})$, $\mathcal{B}(B^- \to D^0 l^\prime\bar{\nu}_{l^\prime})$ and $\mathcal{B}(B^- \to D^0 \tau\bar{\nu}_\tau)$ are in agreement with the HQET prediction, PDG values and the BaBar data within the errors; our predictions on the $\mathcal{B}(\overline{B}^0 \to D^+\tau\bar{\nu}_\tau)$ agrees with the HQET prediction, but is smaller than the value given by the BaBar Collaboration and the PDG average value. We finally get
\begin{eqnarray}
\mathcal{R}(D) = 0.320^{+0.018}_{-0.021}.  \label{rd}
\end{eqnarray}
This value is shown in Fig.\ref{fRD}, where the central value and its uncertainty are indicated by the dashed line and the shaded band, respectively. As a comparison, the experimental data reported by the BaBar Collaboration~\cite{BABAR_Lees:2012xj, BABAR_Lees:2013uzd}, the Belle Collaboration~\cite{BELLE_Huschle:2015rga} and the weighted average of those experimental measurements (HFAG average)~\cite{HFAG_Amhis:2014hma} are presented. The HQET prediction~\cite{HQET_Fajfer:2012vx, HQET_Tanaka:2010se}, the LQCD prediction~\cite{Aoki:2016frl} and the LCSR prediction~\cite{Wang:2017jow} are presented as a comparison.

\section{summary}

In the paper, we have adopted the LCSR approach to calculate the key components of the $B\to D$ semileptonic decays, i.e. the $B\to D$ TFFs. The LCSR predictions on the $B\to D$ TFFs depend heavily on the $D$-meson DAs. At present, we have little knowledge on the $D$-meson twist-3 DAs, and the rough approximation $\phi_{3;D}^{p} \simeq \phi_{2;D}$ is usually adopted. In the paper, we have constructed a new model for the twist-3 DAs $\phi_{3;D}^p$ and $\phi_{3;D}^\sigma$. The input parameters of the twist-3 DAs have been fixed by using the normalization condition, the average value of the $D$-meson transverse momentum and the the moments $\left<\xi_p^n\right>_D$ and $\left<\xi_\sigma^n\right>_D$, which have been calculated by using the QCD SVZ sum rules within the framework of BFT up to NLO level.

Taking $n=0$ in sum rules (\ref{srxips}) and (\ref{srxipt}) and using the normalization conditions $\left<\xi_p^0\right>_D = \left<\xi_\sigma^0\right>_D = 1$, we obtain the sum rules for $\mu_\pi^{p}$ and $\mu_\pi^{\sigma}$, leading to $\mu_D^p = 2.535^{+0.136}_{-0.131} {\rm GeV}$ and $\mu_D^\sigma = 2.534^{+0.267}_{-0.246} {\rm GeV}$ at the scale $\mu = 2 \rm GeV$. The twist-3 DA moments up to $4_{\rm th}$-order, at the scale $\mu = 2 \rm GeV$, are
\begin{eqnarray}
\left<\xi_p^1\right>_D &=& -0.484^{+0.075}_{-0.080},\; \left<\xi_\sigma^1\right>_D = -0.381^{+0.068}_{-0.071}, \\
\left<\xi_p^2\right>_D &=& +0.400^{+0.057}_{-0.052}, \; \left<\xi_\sigma^2\right>_D = +0.296^{+0.037}_{-0.033},  \\
\left<\xi_p^3\right>_D &=& -0.277^{+0.037}_{-0.041}, \; \left<\xi_\sigma^3\right>_D = -0.190^{+0.043}_{-0.044}, \\
\left<\xi_p^4\right>_D &=& +0.242^{+0.035}_{-0.033}, \; \left<\xi_\sigma^4\right>_D = +0.156^{+0.024}_{-0.022}.
\end{eqnarray}
Using the determined $D$-meson twist-3 DAs, we have found that the contributions from the twist-3 DAs are large, which are added up to $39\%$ for LO $f^{B\to D}_+(0)$. We have also shown how various models of the twist-3 DA $\phi_{3;D}^p$ affect the $B\to D$ TFF $f^{B\to D}_{+}(q^2)$. Fig.\ref{ftff1q_comparison} shows that the models $\phi_{3;D}^{p,II,IV}$ bring about $(4-5)\%$ error for the LO $f^{B\to D}_{+}(q^2)$, thus a proper $\phi_{3;D}^{p}$ shall be important for a precise prediction. Figs.(\ref{ftffq_comparison}, \ref{fdgamma}) show that the TFF $f^{B\to D}_{+}(q^2)$ and the differential decay rates for the decay $\overline{B}^0 \to D^+l\bar{\nu}_{l}$ are in agreement with the experimental measurements within errors.

Previous SM theoretical predictions for the ratio $\mathcal{R}(D)$ are always lower than the experimental measurements, some people thus think this inconsistency could indicate a signal of NP. In combination with the LQCD predictions with the LCSR predictions for the TFFs $f^{B\to D}_{+,0}(q^2)$, we achieve a more reliable prediction of the TFFs in whole physical region; and we further predict, $\mathcal{R}(D) = 0.320^{+0.018}_{-0.021}$, whose central value is slightly larger than previous SM predictions and is within $1\sigma$ deviation from the 2015 Belle data. At present the data are still of large errors, our prediction is still about $1.5\sigma$ deviation from the HFAG average of the Belle and BABAR data, we need further accurate measurements of the experiment to confirm whether there is signal of NP from the ratio $\mathcal{R}(D)$.  \\

{\bf Acknowledgments}:
This work was supported in part by the Natural Science Foundation of China under Grant No.11547015, No.11625520 , No.11575110, No.11765007, and No.11605043.

\appendix

\section{Expressions for the terms in the sum rules (\ref{srxips}, \ref{srxipt})}

We present the exprssions for the terms in the sum rules (\ref{srxips},\ref{srxipt}) in the following,
\begin{widetext}
\begin{eqnarray}
{\rm Im} I_{D,\rm pert.}^p &=& \frac{3s}{16\pi(n+1)(n+2)} \left\{ (-1)^n \left[ (n+2) - (n+3)\frac{m_c^2}{s} \right] + \left( 1 - 2\frac{m_c^2}{2} \right)^{n+2} \left[ (n+2) - (n+1)\frac{m_c^2}{s} \right] \right\} \nonumber\\
&+& \frac{3m_c^2}{16\pi(n+1)(n+2)} \left\{ \left[ (2n+3) - 2(n+1)\frac{m_c^2}{s} \right] \left( 1 - 2\frac{m_c^2}{s} \right)^{n+1} + (-1)^n \right\},\\
\hat{L}_M I_{D,\left<\bar{q}q\right>}^p &=& \frac{\left<\bar{q}q\right>}{M^2} \frac{(-1)^n}{2} \exp\left[-\frac{m_c^2}{M^2}\right] \left( m_q \frac{m_c^2}{M^2} - 2m_c + m_q(n+1) \right), \nonumber\\
\hat{L}_M I_{D,\left<G^2\right>}^p &=& \frac{\left<\alpha_sG^2\right>}{24\pi} \left[ 3(n+1)\mathcal{H}(n,0,1,1) + 2n(n^2-1)\mathcal{H}(n-2,1,2,1) - m_c^2 n \mathcal{H}(n,1,0,2) + 3m_c^2 \mathcal{H}(n,0,1,2) \right. \nonumber\\
&+& \left. 2m_c^2n(n-1)\mathcal{H}(n-1,1,2,2) - 2m_c^4\mathcal{H}(n,1,0,3) \right], \\
\hat{L}_M I_{D,\left<\bar{q}Gq\right>}^p &=& \frac{\left<g_s\bar{q}\sigma TGq\right>}{(M^2)^2} \frac{(-1)^n}{36} \exp\left[-\frac{m_c^2}{M^2}\right] \left\{ -4m_q\left(\frac{m_c^2}{M^2}\right)^2 + 3\left[3m_c + m_q(3-4n)\right]\frac{m_c^2}{M^2} + 18(n-1)m_c \right. \nonumber\\
&-& \left. n(8n-5)m_q \right\}, \nonumber\\
\hat{L}_M I_{D,\left<\bar{q}q\right>^2}^p &=& \frac{\left<g_s\bar{q}q\right>^2}{(M^2)^2} \frac{(-1)^n}{81} \exp\left[-\frac{m_c^2}{M^2}\right] \left[ 3\frac{m_c^2}{M^2} + (2n^2+7n-12) \right], \nonumber\\
\hat{L}_M I_{D,\left<G^3\right>}^p &=& \frac{\left<g_s^3fG^3\right>}{2880\pi^2} \left[ -280\mathcal{F}(n,-1,0,2) - 110m_c^2\mathcal{F}(n,-1,0,3) + 200\mathcal{F}(n,-1,1,2) + 140m_c^2\mathcal{F}(n,-1,1,3) \right.\nonumber\\
&-& 45n\mathcal{F}(n,-1,2,2) - 45m_c^2\mathcal{F}(n,-1,2,3) + 210\mathcal{F}(n,-2,0,2) - 280\mathcal{F}(n,-2,1,2) \nonumber\\
&+& 70\mathcal{F}(n,-2,2,2) - 255n\mathcal{F}(n,-2,3,2) - 255m_c^2\mathcal{F}(n,-2,3,3) - 10n(n-1)\mathcal{F}(n-2,-1,3,2) \nonumber\\
&-& 70\left( 3\frac{m_c^2}{M^2} - 2 \right) \widetilde{\mathcal{F}}(n,0,0,2) - 110m_c^2 \widetilde{\mathcal{F}}(n,0,0,3) + 40 \left( 7\frac{m_c^2}{M^2} - 9 \right) \widetilde{\mathcal{F}}(n,0,1,2) \nonumber\\
&+& 140m_c^2 \widetilde{\mathcal{F}}(n,0,1,3) - \left( 70\frac{m_c^2}{M^2} + 45n -140 \right) \widetilde{\mathcal{F}}(n,0,2,2) - 45m_c^2 \widetilde{\mathcal{F}}(n,0,2,3) \nonumber\\
&+& 255n \left( \frac{m_c^2}{M^2} - 2 \right) \widetilde{\mathcal{F}}(n,0,3,2) + 255m_c^2 \left( \frac{m_c^2}{M^2} - 3 \right) \widetilde{\mathcal{F}}(n,0,3,3) - 10n(n-1) \widetilde{\mathcal{F}}(n-2,0,3,2) \nonumber\\
&+& 210 \widetilde{\mathcal{F}}(n,1,0,2) - 180 \widetilde{\mathcal{F}}(n,1,1,2) + 70 \widetilde{\mathcal{F}}(n,1,2,2) - 255n \widetilde{\mathcal{F}}(n,1,3,2) - 255m_c^2 \widetilde{\mathcal{F}}(n,1,3,3) \nonumber\\
&+& 70\mathcal{H}(n,0,0,2) + 80m_c^2\mathcal{H}(n,0,0,3) + 210m_c^4\mathcal{H}(n,0,0,4) - 45n\mathcal{H}(n,0,1,2) - 30(4n-1)m_c^2 \mathcal{H}(n,0,1,3) \nonumber\\
&-& 360m_c^4\mathcal{H}(n,0,1,4) - 255n\mathcal{H}(n,1,0,2) - 30m_c^2(n+16)\mathcal{H}(n,1,0,3) + 18m_c^4(2n-9)\mathcal{H}(n,1,0,4) \nonumber\\
&+& 144m_c^6\mathcal{H}(n,1,0,5) - 30n(n-1)\mathcal{H}(n-2,0,2,2) - 10n(n-1)\mathcal{H}(n-2,1,1,2) \nonumber\\
&-& 20m_c^2n(n-1)\mathcal{H}(n-2,1,1,3) - 20n^2(n^2-1)\mathcal{H}(n-2,1,3,2) - 80m_c^2n^2(n-1)\mathcal{H}(n-2,1,3,3) \nonumber\\
&-& 120m_c^4n(n-1)\mathcal{H}(n-2,1,3,4) + \frac{5(-1)^n}{2(M^2)^2} \exp \left[-\frac{m_c^2}{M^2}\right] \left\{ \left( \ln\frac{M^2}{\mu^2} - \gamma_E \right) \left[ \left( 306\frac{m_c^2}{M^2} + 306n + 56 \right) \delta^{n0} \right.\right. \nonumber\\
&-& 4n(n-1)\theta(n-2) + \left( 102(2n+3) \frac{m_c^2}{M^2} + 204n^2 + 306n + 56 \right) \theta(n-1) + 102 \left(\frac{m_c^2}{M^2}\right)^2 \nonumber\\
&+& \left. 6(17n-52)\frac{m_c^2}{M^2} - 222n - 32 \right] + \left( 459\frac{m_c^2}{M^2} + 306n - 28 \right) \delta^{n0} - 4n(n-1) \theta(n-2) + \left( 153(2n+3)\frac{m_c^2}{M^2} \right. \nonumber\\
&+& \left.\left.\left. 204n^2 + 306n + 56 \right) \theta(n-1) + 187\left(\frac{m_c^2}{M^2}\right)^2 + 9(17n-52)\frac{m_c^2}{M^2} - 222n - 32 \right\} \right], \\
{\rm Im} I^\sigma_{D,\rm pert.} &=& \frac{3s}{16\pi(n+2)(n+3)} \left\{ (-1)^n \left[ (n+2) - (n+3)\frac{m_c^2}{s} \right] + \left( 1 - 2\frac{m_c^2}{2} \right)^{n+2} \left[ (n+2) - (n+1)\frac{m_c^2}{s} \right] \right\}, \\
\hat{L}_M I_{D,\left<\bar{q}q\right>}^\sigma &=& \frac{\left<\bar{q}q\right>}{M^2} (n+1) \frac{(-1)^n}{2} \exp\left[-\frac{m_c^2}{M^2}\right] m_q, \nonumber\\
\hat{L}_M I_{D,\left<G^2\right>}^\sigma &=& \frac{\left<\alpha_sG^2\right>}{24\pi}(n+1) \left[ \mathcal{H}(n,0,1,1) + 2n(n-1)\mathcal{H}(n-2,1,2,1) - m_c^2\mathcal{H}(n,1,0,2) \right],\\
\hat{L}_M I_{D,\left<\bar{q}Gq\right>}^\sigma &=& \frac{\left<g_s\bar{q}\sigma TGq\right>}{(M^2)^2} (n+1) \frac{(-1)^n}{36} \exp\left[-\frac{m_c^2}{M^2}\right] \left[ -4m_q\frac{m_c^2}{M^2} - \left[6m_c+(8n+1)m_q\right] \right], \nonumber\\
\hat{L}_M I_{D,\left<\bar{q}q\right>^2}^\sigma &=& \frac{\left<g_s\bar{q}q\right>^2}{(M^2)^2} (n+1) \frac{(-1)^n}{81} \exp\left[-\frac{m_c^2}{M^2}\right] \left( 2n-5 - 2\frac{m_c^2}{M^2} \right), \nonumber\\
\hat{L}_M I_{D,\left<G^3\right>}^\sigma &=& \frac{\left<g_s^3fG^3\right>}{2880\pi^2}(n+1) \left[ 15\mathcal{F}(n,-1,2,2) - 10(n+1)(3n-1)\mathcal{F}(n,-1,3,2) - 10m_c^2(3n-1)\mathcal{F}(n,-1,3,3) \right. \nonumber\\
&-& 255\mathcal{F}(n,-2,3,2) - 5n(3n+5)\mathcal{F}(n-1,-1,3,2) - 30n m_c^2\mathcal{F}(n-1,-1,3,3) \nonumber\\
&+& 15n(n-1)\mathcal{F}(n-2,-1,3,2) + 15 \widetilde{\mathcal{F}}(n,0,2,2) + 5\left( 51\frac{m_c^2}{M^2} - 2(3n^2+2n+50) \right) \widetilde{\mathcal{F}}(n,0,3,2) \nonumber\\
&-& 10(3n-1) m_c^2 \widetilde{\mathcal{F}}(n,0,3,3) - 5n(3n+5) \widetilde{\mathcal{F}}(n-1,0,3,2) - 30n m_c^2 \widetilde{\mathcal{F}}(n-1,0,3,3) \nonumber\\
&+& 15n(n-1) \widetilde{\mathcal{F}}(n-2,0,3,2) - 255 \widetilde{\mathcal{F}}(n,1,3,2) + 15 \mathcal{H}(n,0,1,2) + 20m_c^2\mathcal{H}(n,0,1,3) \nonumber\\
&-& 20(n+1)\mathcal{H}(n,0,2,2) - 40m_c^2\mathcal{H}(n,0,2,3) - 255\mathcal{H}(n,1,0,2) - 30m_c^2\mathcal{H}(n,1,0,3) + 36m_c^4\mathcal{H}(n,1,0,4) \nonumber\\
&+& 90(n+1)\mathcal{H}(n,1,1,2) - 20m_c^2(n-9)\mathcal{H}(n,1,1,3) - 180m_c^4\mathcal{H}(n,1,1,4) + 240m_c^4\mathcal{H}(n,1,2,4) \nonumber\\
&-& 40(n+1)(n+2)\mathcal{H}(n,2,1,2) - 160(n+1)m_c^2\mathcal{H}(n,2,1,3) + 60n(n+1)\mathcal{H}(n-1,0,3,2) \nonumber\\
&+& 120n m_c^2\mathcal{H}(n-1,0,3,3) + 5n(7n+5)\mathcal{H}(n-1,1,1,2) + 80n m_c^2\mathcal{H}(n-1,1,1,3)  \nonumber\\
&-& 60n(n+1)\mathcal{H}(n-1,1,2,2) - 160n m_c^2\mathcal{H}(n-1,1,2,3) + 40n m_c^2\mathcal{H}(n-1,2,1,3) \nonumber\\
&-& 30n(n-1)\mathcal{H}(n-2,0,3,2) + 5n(n-1)\mathcal{H}(n-2,1,1,2) - 30n(n-1)\mathcal{H}(n-2,1,2,2) \nonumber\\
&-& 20n(n^2-1)\mathcal{H}(n-2,1,3,2) - 40n(n-1)m_c^2\mathcal{H}(n-2,1,3,3) + \frac{5(-1)^n}{2(M^2)^2} \exp \left[-\frac{m_c^2}{M^2}\right] \nonumber\\
&\times& \left\{ \left( \ln\frac{M^2}{\mu^2} - \gamma_E + 1 \right ) \left[ 306 \delta^{n0} + 6n(n-1) \theta(n-2) + \left( 12n \frac{m_c^2}{M^2} + 6n^2 + 214n + 306 \right) \theta(n-1) \right.\right. \nonumber\\
&-& \left.\left.\left. 2(6n-53)\frac{m_c^2}{M^2} - 2(6n^2+4n+97) \right] + 6n\frac{m_c^2}{M^2} \theta(n-1) - (6n-53)\frac{m_c^2}{M^2} \right\} \right],
\end{eqnarray}
where
\begin{eqnarray}
\mathcal{F}(n,a,b,c) &=& \frac{1}{\left(M^2\right)^c} \exp \left[ -\frac{m_c^2}{M^2} \right] \sum^n_{k=0} \frac{(-1)^k n! \Gamma(k+b+1)}{k! (n-k)!} \times \sum^\infty_{l=-a} \frac{\Gamma(l+c) \Gamma(n-k+l+a+1)}{\Gamma(n+l+a+b+2)} \nonumber\\
&\times& \sum^l_{i=0} \frac{1}{i! (l-i)! (l-i+c-1)!} \left( -\frac{m_c^2}{M^2} \right)^{l-i}, \nonumber\\
\widetilde{\mathcal{F}}(n,i,b,c) &=& \frac{1}{\left(M^2\right)^c} \exp \left[-\frac{m_c^2}{M^2}\right] \sum^{n-1-i}_{k=0} \frac{(-1)^k n! \Gamma(k+b+1) \Gamma(n+b+i) \Gamma(n-k-i)}{k! (n-k)! \Gamma(n+b) \Gamma(n+b+1)} \nonumber\\
\mathcal{H}(n,a,b,c) &=& \frac{1}{(c-1)!} \frac{1}{(M^2)^c} \int^1_0 dx (2x-1)^n x^a (1-x)^{b-c} \exp \left[ -\frac{m_c^2}{M^2(1-x)} \right]. \nonumber
\end{eqnarray}
\end{widetext}


\begin{thebibliography}{s2}

\bibitem{BABAR_Lees:2012xj} J.~P.~Lees {\it et al.} [BaBar Collaboration], ``Evidence for an excess of $\bar{B} \to D^{(\ast)} \tau^-\bar{\nu}_\tau$ decays,'' Phys.\ Rev.\ Lett.\  {\bf 109}, 101802 (2012).

\bibitem{BABAR_Lees:2013uzd} J.~P.~Lees {\it et al.} [BaBar Collaboration], ``Measurement of an Excess of $\bar{B} \to D^{(\ast)}\tau^- \bar{\nu}_\tau$ Decays and Implications for Charged Higgs Bosons,'' Phys.\ Rev.\ D {\bf 88}, 072012 (2013).

\bibitem{BELLE_Huschle:2015rga} M.~Huschle {\it et al.} [Belle Collaboration], ``Measurement of the branching ratio of $\bar{B} \to D^{(\ast)} \tau^- \bar{\nu}_\tau$ relative to $\bar{B} \to D^{(\ast)} \ell^- \bar{\nu}_\ell$ decays with hadronic tagging at Belle,'' Phys.\ Rev.\ D {\bf 92}, 072014 (2015).

\bibitem{HFAG_Amhis:2014hma} Y.~Amhis {\it et al.} [Heavy Flavor Averaging Group (HFAG)], ``Averages of $b$-hadron, $c$-hadron, and $\tau$-lepton properties as of summer 2014,'' arXiv:1412.7515 [hep-ex]. For the update of the average of $\mathcal{R}(D)$ one can see the web page: http://www.slac.stanford.edu/xorg/hfag/semi/index.html.

\bibitem{HQET_Fajfer:2012vx} S.~Fajfer, J.~F.~Kamenik and I.~Nisandzic, ``On the $B \to D^\ast \tau \bar{\nu}_{\tau}$ Sensitivity to New Physics,'' Phys.\ Rev.\ D {\bf 85}, 094025 (2012).

\bibitem{HQET_Tanaka:2010se} M.~Tanaka and R.~Watanabe, ``Tau longitudinal polarization in $\bar{B}\to D \tau \bar{\nu}$ and its role in the search for charged Higgs boson,'' Phys.\ Rev.\ D {\bf 82}, 034027 (2010).

\bibitem{LQCD_Lattice:2015rga} J.~A.~Bailey {\it et al.} [MILC Collaboration], ``$B\to Dl\nu$ form factors at nonzero recoil and $|V_{cb}|$ from 2+1-flavor lattice QCD,'' Phys.\ Rev.\ D {\bf 92}, 034506 (2015).

\bibitem{LQCD_Na:2015kha} H.~Na {\it et al.} [HPQCD Collaboration], ``$B \to D l \nu$ form factors at nonzero recoil and extraction of $|V_{cb}|$,'' Phys.\ Rev.\ D {\bf 92}, 054510 (2015); Erratum: [Phys.\ Rev.\ D {\bf 93}, 119906 (2016)].

\bibitem{Aoki:2016frl} S.~Aoki {\it et al.}, ``Review of lattice results concerning low-energy particle physics,'' Eur.\ Phys.\ J.\ C {\bf 77}, 112 (2017).

\bibitem{Bigi:2016mdz} D.~Bigi and P.~Gambino, ``Revisiting $B\to D \ell \nu$,'' Phys.\ Rev.\ D {\bf 94}, 094008 (2016).

\bibitem{NP_Celis:2012dk} A.~Celis, M.~Jung, X.~Q.~Li and A.~Pich, ``Sensitivity to charged scalars in $B\to D^{(\ast)}\tau\nu_\tau$ and $B\to\tau\nu_\tau$ decays,'' JHEP {\bf 1301}, 054 (2013).

\bibitem{NP_Celis:2013jha} A.~Celis, M.~Jung, X.~Q.~Li and A.~Pich, ``$B\to D^{(\ast)}\tau\nu_\tau$ decays in two-Higgs-doublet models,'' J.\ Phys.\ Conf.\ Ser.\  {\bf 447}, 012058 (2013).

\bibitem{NP_Li:2016vvp} X.~Q.~Li, Y.~D.~Yang and X.~Zhang, ``Revisiting the one leptoquark solution to the $R(D^{(\ast)})$ anomalies and its phenomenological implications,'' JHEP {\bf 1608}, 054 (2016).

\bibitem{PQCD_Fan:2015kna} Y.~Y.~Fan, Z.~J.~Xiao, R.~M.~Wang and B.~Z.~Li, ``The $B\to D^{(\ast)} l\nu_l$ decays in the pQCD approach with the Lattice QCD input,'' Chin.\ Sci.\ Bull.\ 60, 2009 (2015).

\bibitem{PQCD_Fan:2013qz} Y.~Y.~Fan, W.~F.~Wang, S.~Cheng and Z.~J.~Xiao, ``Semileptonic decays $B \to D^{(\ast)} l\nu$ in the perturbative QCD factorization approach,'' Chin.\ Sci.\ Bull.\  {\bf 59}, 125 (2014).

\bibitem{LCSR_Zuo:2006dk} F.~Zuo, Z.~H.~Li and T.~Huang, ``Form Factor for $B\to D l \tilde{\nu}$ in Light-Cone Sum Rules With Chiral Current Correlator,'' Phys.\ Lett.\ B {\bf 641}, 177 (2006).

\bibitem{Zuo:2006re} F.~Zuo and T.~Huang, ``$B_c(B)\to Dl \tilde{\nu}$ form-factors in light-cone sum rules and the $D$ meson distribution amplitude,'' Chin.\ Phys.\ Lett.\ {\bf 24}, 61 (2007).

\bibitem{Fu:2013wqa} H.~B.~Fu, X.~G.~Wu, H.~Y.~Han, Y.~Ma and T.~Zhong, ``$|V_{cb}|$ from the semileptonic decay $B\to D \ell \bar{\nu}_{\ell}$ and the properties of the $D$ meson distribution amplitude,'' Nucl.\ Phys.\ B {\bf 884}, 172 (2014).

\bibitem{Zhang:2017rwz} Y.~Zhang, T.~Zhong, X.~G.~Wu, K.~Li, H.~B.~Fu and T.~Huang, ``Uncertainties of the $B\to D$ transition form factor from the $D$-meson leading-twist distribution amplitude,'' Eur.\ Phys.\ J.\ C {\bf 78}, 76 (2018)

\bibitem{Wang:2017jow} Y.~M.~Wang, Y.~B.~Wei, Y.~L.~Shen and C.~D.~L¨¹, ``Perturbative corrections to $B\to D$ form factors in QCD,'' JHEP {\bf 1706}, 062 (2017).

\bibitem{MODELI_Kurimoto:2002sb} T.~Kurimoto, H.~n.~Li and A.~I.~Sanda, ``$B\to D^{(\ast)}$ form-factors in perturbative QCD,'' Phys.\ Rev.\ D {\bf 67}, 054028 (2003).

\bibitem{MODELI_Keum:2003js} Y.~Y.~Keum, T.~Kurimoto, H.~N.~Li, C.~D.~L\"{u} and A.~I.~Sanda, ``Nonfactorizable contributions to $B\to D^{(\ast)}M$ decays,'' Phys.\ Rev.\ D {\bf 69}, 094018 (2004).

\bibitem{MODELII_Li:2008ts} R.~H.~Li, C.~D.~L\"{u} and H.~Zou, ``The $B(B_s)\to D_{(s)}P, D_{(s)}V, D^\ast_{(s)}P$ and $D^\ast_{(s)}V$ decays in the perturbative QCD approach,'' Phys.\ Rev.\ D {\bf 78}, 014018 (2008).

\bibitem{MODELIII_Li:1999kna} H.~n.~Li and B.~Melic, ``Determination of heavy meson wave functions from $B$ decays,'' Eur.\ Phys.\ J.\ C {\bf 11}, 695 (1999).

\bibitem{MOLELIV_Guo:1991eb} X.~H.~Guo and T.~Huang, ``Hadronic wave functions in $D$ and $B$ decays,'' Phys.\ Rev.\ D {\bf 43}, 2931 (1991).

\bibitem{MODELV_Grozin:1996pq} A.~G.~Grozin and M.~Neubert, ``Asymptotics of heavy meson form-factors,'' Phys.\ Rev.\ D {\bf 55}, 272 (1997).

\bibitem{MODELVI_Kawamura:2001jm} H.~Kawamura, J.~Kodaira, C.~F.~Qiao and K.~Tanaka, ``$B$-meson light cone distribution amplitudes in the heavy quark limit,'' Phys.\ Lett.\ B {\bf 523}, 111 (2001); Erratum: [Phys.\ Lett.\ B {\bf 536}, 344 (2002)].

\bibitem{SVZ_Shifman:1978bx} M.~A.~Shifman, A.~I.~Vainshtein and V.~I.~Zakharov, ``QCD and Resonance Physics. Theoretical Foundations,'' Nucl.\ Phys.\ B {\bf 147}, 385 (1979).

\bibitem{Huang:1986wm} T.~Huang, X.~n.~Wang, X.~d.~Xiang and S.~J.~Brodsky, ``The Quark Mass and Spin Effects in the Mesonic Structure,'' Phys.\ Rev.\ D {\bf 35}, 1013 (1987).

\bibitem{Huang:1989gv} T.~Huang and Z.~Huang, ``Quantum Chromodynamics in Background Fields,'' Phys.\ Rev.\ D {\bf 39}, 1213 (1989).

\bibitem{Zhong:2014jla} T.~Zhong, X.~G.~Wu, Z.~G.~Wang, T.~Huang, H.~B.~Fu and H.~Y.~Han, ``Revisiting the Pion Leading-Twist Distribution Amplitude within the QCD Background Field Theory,'' Phys.\ Rev.\ D {\bf 90}, 016004 (2014).

\bibitem{Kurimoto:2002sb} T.~Kurimoto, H.~n.~Li and A.~I.~Sanda, ``$B\to D^{(\ast)}$ form-factors in perturbative QCD,'' Phys.\ Rev.\ D {\bf 67}, 054028 (2003).

\bibitem{Duplancic:2008ix} G.~Duplancic, A.~Khodjamirian, T.~Mannel, B.~Melic and N.~Offen, ``Light-cone sum rules for $B\to \pi$ form factors revisited,'' JHEP {\bf 0804}, 014 (2008).

\bibitem{BHL1} S. J. Brodsky, T. Huang, and G. P. Lepage, in \textit{Particles and Fields-2}, Proceedings of the Banff Summer Institute, Banff; Alberta, 1981, edited by A. Z. Capri and A. N. Kamal (Plenum, New York, 1983), p. 143;

\bibitem{BHL2} G. P. Lepage, S. J. Brodsky, T. Huang, and P. B.Mackenize, in \textit{Particles and Fields-2}, Proceedings of the Banff Summer Institute, Banff; Alberta, 1981, edited by A. Z. Capri and A. N. Kamal (Plenum, New York, 1983), p. 83;

\bibitem{BHL3} T. Huang, in \textit{Proceedings ofXXth International Conference on High Energy Physics}, Madison, Wisconsin, 1980, edited by L. Durand and L. G Pondrom, AIP Conf. Proc. No. 69 (AIP, New York, 1981), p. 1000.

\bibitem{Lepage:1980fj} G.~P.~Lepage and S.~J.~Brodsky, ``Exclusive Processes in Perturbative Quantum Chromodynamics,'' Phys.\ Rev.\ D {\bf 22}, 2157 (1980).

\bibitem{PDG_Olive:2016xmw} C.~Patrignani {\it et al.} [Particle Data Group], ``Review of Particle Physics,'' Chin.\ Phys.\ C {\bf 40}, 100001 (2016)  and 2017 update.

\bibitem{SRREV_Colangelo:2000dp} P.~Colangelo and A.~Khodjamirian, ``QCD sum rules, a modern perspective,'' In *Shifman, M. (ed.): At the frontier of particle physics, vol. 3* 1495-1576, [hep-ph/0010175].

\bibitem{RGE_Yang:1993bp} K.~C.~Yang, W.~Y.~P.~Hwang, E.~M.~Henley and L.~S.~Kisslinger, ``QCD sum rules and neutron proton mass difference,'' Phys.\ Rev.\ D {\bf 47}, 3001 (1993).

\bibitem{RGE_Hwang:1994vp} W.~Y.~P.~Hwang and K.~C.~Yang, ``QCD sum rules: $\Delta-N$ and $\Sigma_0-\Lambda$ mass splittings,'' Phys.\ Rev.\ D {\bf 49}, 460 (1994).

\bibitem{Huang:2004tp} T.~Huang, X.~H.~Wu and M.~Z.~Zhou, ``Twist three distribute amplitudes of the pion in QCD sum rules,'' Phys.\ Rev.\ D {\bf 70}, 014013 (2004).

\bibitem{Huang:2005av} T.~Huang, M.~Z.~Zhou and X.~H.~Wu, ``Twist-3 distribution amplitudes of the pion and kaon from the QCD sum rules,'' Eur.\ Phys.\ J.\ C {\bf 42}, 271 (2005).

\bibitem{Braun:1989iv} V.~M.~Braun and I.~E.~Filyanov, ``Conformal Invariance and Pion Wave Functions of Nonleading Twist,'' Z.\ Phys.\ C {\bf 48}, 239 (1990).

\bibitem{Ball:1998sk} P.~Ball, V.~M.~Braun, Y.~Koike and K.~Tanaka, ``Higher twist distribution amplitudes of vector mesons in QCD: Formalism and twist - three distributions,'' Nucl.\ Phys.\ B {\bf 529}, 323 (1998).

\bibitem{Ball:1998je} P.~Ball, ``Theoretical update of pseudoscalar meson distribution amplitudes of higher twist: The Nonsinglet case,'' JHEP {\bf 9901}, 010 (1999).

\bibitem{Zhong:2016kuv} T.~Zhong, X.~G.~Wu, T.~Huang and H.~B.~Fu, ``Heavy Pseudoscalar Twist-3 Distribution Amplitudes within QCD Theory in Background Fields,'' Eur.\ Phys.\ J.\ C {\bf 76}, 509 (2016).

\bibitem{Huang:2004su} T.~Huang and X.~G.~Wu, ``A Model for the twist-3 wave function of the pion and its contribution to the pion form-factor,''  Phys.\ Rev.\ D {\bf 70}, 093013 (2004).

\bibitem{Huang:2004hw} T.~Huang and X.~G.~Wu, ``Consistent calculation of the B to pi transition form-factor in the whole physical region,'' Phys.\ Rev.\ D {\bf 71}, 034018 (2005).

\bibitem{Wang:2008xt} W.~Wang, Y.~L.~Shen and C.~D.~L\"{u}, ``Covariant Light-Front Approach for $B_c$ transition form factors,'' Phys.\ Rev.\ D {\bf 79}, 054012 (2009).

\bibitem{Glattauer:2015teq} R.~Glattauer {\it et al.} [Belle Collaboration], ``Measurement of the decay $B\to D\ell\nu_\ell$ in fully reconstructed events and determination of the Cabibbo-Kobayashi-Maskawa matrix element $|V_{cb}|$,'' Phys.\ Rev.\ D {\bf 93}, 032006 (2016).

\bibitem{Aubert:2008yv} B.~Aubert {\it et al.} [BaBar Collaboration], ``Measurements of the Semileptonic Decays $\bar{B}\to Dl\bar{\nu}$ and $\bar{B}\to D^\ast l\bar{\nu}$ Using a Global Fit to $DXl\bar{\nu}$ Final States,'' Phys.\ Rev.\ D {\bf 79}, 012002 (2009).

\bibitem{Aubert:2009ac} B.~Aubert {\it et al.} [BaBar Collaboration], ``Measurement of $|V_{cb}|$ and the Form-Factor Slope in $\bar{B}\to D l^- \bar{\nu}_l$ Decays in Events Tagged by a Fully Reconstructed B Meson,'' Phys.\ Rev.\ Lett.\  {\bf 104}, 011802 (2010)


\end{thebibliography}
\end{document}